\documentclass[aps,prl,reprint,groupedaddress]{revtex4-1}

\usepackage{graphicx}
\usepackage{hyperref}

\begin{document}

\title{Temperature-dependent magnetocrystalline anisotropy of 
rare earth/transition metal permanent magnets from first principles:
the light RCo\textsubscript{5} (R=Y, La--Gd) intermetallics}

\author{Christopher E.\ Patrick}
\email{c.patrick.1@warwick.ac.uk}
\author{Julie B.\ Staunton}
\affiliation{Department of Physics, University of Warwick, Coventry CV4 7AL, United Kingdom}

\date{\today}

\begin{abstract}
Computational design of more efficient rare earth/transition metal (RE-TM) permanent magnets 
requires accurately calculating the magnetocrystalline anisotropy (MCA) at finite 
temperature, since this property places an upper bound on the 
coercivity.
Here, we present a first-principles methodology to calculate the MCA of RE-TM magnets which 
fully accounts for the effects of temperature on the underlying electrons. 
The itinerant electron TM magnetism is described within the disordered
local moment picture, and the localized RE-4\emph{f}  magnetism
is described within crystal field theory.
We use our model, which is free of adjustable parameters, to calculate the MCA of the RECo\textsubscript{5} 
(RE = Y, La--Gd) magnet family for temperatures 0--600~K.
We correctly find a huge uniaxial anisotropy for SmCo\textsubscript{5} 
(21.3~MJm\textsuperscript{-3} at 300~K)
and two finite temperature spin reorientation transitions
for NdCo\textsubscript{5}.
The calculations also demonstrate dramatic valency effects
in CeCo\textsubscript{5} and PrCo\textsubscript{5}.
Our calculations provide quantitative, first-principles insight into
several decades of RE-TM experimental studies.
\end{abstract}

\maketitle

The excellent properties of rare earth/transition metal (RE-TM) permanent magnets 
have facilitated a number of technological revolutions in the last 50 years.
Now, the urgent need for a low carbon,
low emission economy is driving a global research effort dedicated to
improving RE-TM performance for more efficient deployment
in the drive motors of hybrid and electric vehicles~\cite{Nakamura2017}.
RE-TM magnets 
combine the large volume magnetization and high Curie 
temperature of the elemental TM magnets Fe or Co with the potentially 
huge magnetocrystalline anisotropy (MCA) of the REs~\cite{Coey2011}.
The REs consist of Sc, Y and the lanthanides La--Lu, but 
it is Y and the light lanthanides, i.e.\ those with smaller atomic
masses than Gd, which are most attractive for applications due to their lower criticality~\cite{Gutfleisch2011}.
Nd and Sm stand out thanks to the highly successful 
Nd-Fe-B and Sm-Co magnets~\cite{Sagawa1984,Croat1984,Strnat1972,Strnat1967},
but Ce has the attraction of having a low cost and high abundance 
compared to the other REs~\cite{Pathak2015}.

Traditionally, RE-TM magnet research has been driven by experiments.
First-principles computational modelling can uncover fundamental
physical principles and provide new directions for RE-TM magnet design~\cite{Vekilova2019,Tatetsu2018}, 
but faces two challenges.
First, RE-TM magnetism originates from both itinerant electrons
and more localized lanthanide 4$f$ electrons~\cite{Perdew1981}.
Although the local spin-density approximation to density-functional theory~\cite{Kohn1965} (DFT)
satisfactorily describes the itinerant electrons, the 4$f$ electrons
require specialist techniques like dynamical mean field theory~\cite{Locht2016,Delange2017}, 
the local self-interaction correction (LSIC)~\cite{Lueders2005},
the open-core approximation~\cite{Brooks19912,Soderlind2017,Fukazawa2017}, or 
Hubbard $+U$ models~\cite{Waller2016,Larson2004}.
Second, DFT calculations are most easily performed at zero temperature,
but under actual operating conditions the RE-TM magnetic moments are subject
to a considerable level of thermal disorder~\cite{Gyorffy1985}.
The disordered local moment (DLM) picture accounts for this disorder within DFT~\cite{Gyorffy1985} and,
combined with
the LSIC has been 
used successfully to calculate magnetizations, Curie temperatures and phase diagrams
of itinerant electron and RE-based magnets~\cite{Hughes2007,MendiveTapia2017,Patrick20182}.
DFT-DLM studies of the MCA have also been performed on itinerant electron and Gd-based magnets~\cite{Staunton2004,Matsumoto2014,Patrick2018},
but it is important to realise that these materials are special cases, where
there is no contribution to the MCA from 4$f$ electrons interacting with
their local environment (the crystal field).
A first-principles, finite temperature theory which accounts for these crystal field effects
(and is therefore applicable to general RE-TM magnets
like Nd-Fe-B or Sm-Co) has, up to now, proven elusive.

\begin{figure}
\includegraphics{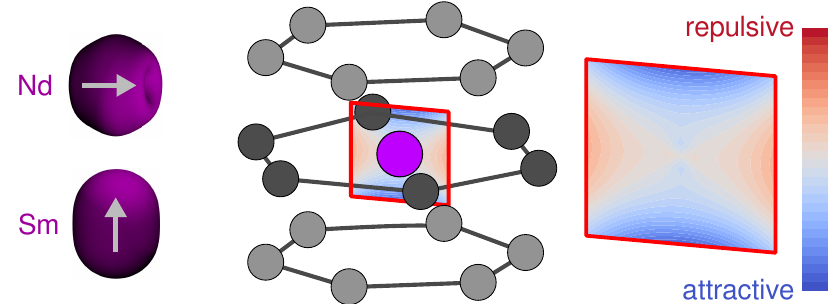}
\caption{
Preferred orientation of RE-4$f$ charge clouds (with magnetic moments
indicated by arrows) for Nd and Sm in the RECo\textsubscript{5} crystal field,
shown on the contour plot.
The Co and RE atoms are shaded in grey and purple, respectively.
Repulsive (attractive) corresponds to high (low) potential energy.
\label{fig.1}}
\end{figure}
In this Rapid Communication, we rectify this situation and demonstrate a 
first-principles theory of the MCA of RE-TM magnets including the crystal field interaction.
Fundamentally, the theory takes  
a model originally developed by experimentalists, and obtains
all of the quantities required by the model from DFT-based calculations.
We demonstrate the theory on the RECo\textsubscript{5} family of magnets,
(RE = Y, La, Ce, Pr, Nd, Sm and Gd).
The RECo\textsubscript{5} phase, shown in Fig.~\ref{fig.1},
is important due to its presence
in SmCo\textsubscript{5} and in the cell-boundary phase of commercial
Sm\textsubscript{2}Co\textsubscript{17}~\cite{Wernick1959,Strnat1967,Kumar1988}.
We calculate
anisotropy constants and spin reorientation transition temperatures
to analyse experimental data obtained 40 years ago~\cite{Ermolenko1976,Klein1975}.

The model partitions the RE-TM magnet into an
itinerant electron subsystem (originating from the TM, and RE valence electrons),
and a subsystem of strongly-localized RE-4$f$
electrons~\cite{Kuzmin2008}.
Critically, the RE ions tend to adopt a 3+ state with a common $s^2d$ valence
structure~\cite{Patrick20192}.
As a consequence, for each RE-TM magnet class
the itinerant electron subsystem is essentially independent of the specific RE~\cite{Kuzmin2008},
so its properties can be obtained for the most computationally convenient prototype (e.g.\ RE = Y or Gd)
The itinerant electrons drive the overall magnetic order, primarily 
determining the Curie temperature 
$T_\mathrm{C}$; for example, $T_\mathrm{C}$   differs by only 20~K 
in Y\textsubscript{2}Fe\textsubscript{14}B and
Nd\textsubscript{2}Fe\textsubscript{14}B~\cite{Herbst1991}.
The itinerant electrons also drive the RE-4$f$ magnetic ordering
through an antiferromagnetic exchange interaction~\cite{Brooks1989}, with
an exchange field of a few hundred Tesla
at cryogenic temperatures~\cite{Loewenhaupt1994}.
RE-RE interactions are relatively weak~\cite{Patrick2017}.

The RE-4$f$ subsystem contributes to the magnetic moment and can have a small
effect on $T_\mathrm{C}$~\cite{Patrick20182}, but its most important contribution
is to the MCA, which in turn provides
an intrinsic mechanism for coercivity~\cite{Brown1957}.
The origin of the potentially huge MCA of RE-TM magnets is 
illustrated in Fig.~\ref{fig.1}.
The itinerant electrons and surrounding ions set up a (primarily) electrostatic
potential with the symmetry of the RE site~\cite{Kuzmin2008},
known as the crystal field (CF).
The CF calculated for RECo\textsubscript{5} is shown as a contour plot in Fig.~\ref{fig.1},
with electrons attracted to the blue regions and repelled
by the red.
Meanwhile the unfilled RE-4$f$ shells form non-spherically-symmetric charge clouds 
coupled to the magnetic moment direction through a strong spin-orbit interaction~\cite{Kuzmin2008}.
The charge clouds are elongated either parallel or perpendicular to the moment direction
depending on Hund's rules, with the opposing examples of Sm and Nd shown in Fig.~\ref{fig.1}.
Placed in the RECo\textsubscript{5} CF, the charge clouds
will preferentially orientate with their elongated part lying in the attractive region, 
generating the MCA~\cite{Newman1989}.
A secondary contribution to the MCA is provided by the itinerant electrons,
with YCo\textsubscript{5}  (which has no RE-4$f$ electrons)
having a MCA energy of  5~MJm$^{-3}$ at
room temperature~\cite{Ermolenko1976},
ten times larger than hexagonal Co~\cite{Sucksmith1954}.

The above ideas are formulated mathematically by introducing
a Hamiltonian for the RE-4$f$ electrons 
$\hat{\mathcal{H}}$~\cite{Tiesong1991}:
\begin{equation}
\hat{\mathcal{H}} = \lambda \hat{\bf L}\cdot \hat{\bf S} +  \mu_0\mu_B (\hat{\bf L} + 2\hat{\bf S}) \cdot {\bf H} 
          + 2\mu_B\hat{\bf S}\cdot{\bf B_\mathrm{exch}}+ \sum_iV({\bf r_i}).
\label{eq.ham}
\end{equation}
Here, $\hat{\bf L}$ and $\hat{\bf S}$ are the orbital and spin angular momentum operators,
where for each RE$^{3+}$ ion $L$ and $S$ are fixed by Hund's rules.
$\lambda$ quantifies the spin-orbit interaction,
and $\mu_B$ is the Bohr magneton.
${\bf H}$ is an external magnetic field,
and ${\bf B_\mathrm{exch}}$ is the exchange field originating from the 
itinerant electrons.
$V({\bf r_i})$ is the crystal field potential, where $i$ labels each $4f$ electron.
The Hamiltonian in Eq.~\ref{eq.ham} acts upon the RE-4$f$ wavefunction which can
be expressed schematically as a radial part multiplied by the angular part
$|J,L,S,M_J\rangle$, where the quantities within the ket are standard quantum
numbers~\cite{Kuzmin2008}.
Equation~\ref{eq.ham} is diagonalized within the manifold of states
$|J,L,S,M_J=J,J-1,J-2,...,-|J|\rangle$.
We consider the ground $J = |L-S|$ multiplet, along with the first
excited multiplet $J = |L-S|+1$ for Pr and Nd, and also the
next excited multiplet $J = |L-S|+2$ for Sm.
Angular parts of the matrix elements of the CF term in Eq.~\ref{eq.ham} are calculated
by decomposing the states into $|L,S,M_L,M_S\rangle$ form,
and then using the operator form of the potential~\cite{Stevens1952}, which
introduces Clebsch-Gordan coefficients and $l$-dependent 
(Stevens) prefactors~\cite{Kuzmin2008,Chilton2013}.
The radial parts are incorporated into the CF coefficients $B_{lm}$~\cite{Patrick20192},
described in more detail below.
For a given temperature $T$, we use the eigenvalue spectrum of 
$\hat{\mathcal{H}}$ to construct 
the partition function $Z_\mathrm{RE}$
and the RE-4$f$ free energy contribution 
$ F_\mathrm{RE}(T,{\bf B_\mathrm{exch}},{\bf H}) = -k_\mathrm{B}T \ln Z_\mathrm{RE}$.

The quantities forming the itinerant contribution to the free energy $F_\mathrm{itin}$ are determined from DFT-DLM calculations.
$F_\mathrm{itin}$ depends on the Co magnetization ${\bf M_\mathrm{Co}}$:
\begin{eqnarray}
F_\mathrm{itin}(T,{\bf \hat{M}_\mathrm{Co}},{\bf H}) &=& K_1\sin^2\theta + K_2\sin^4\theta  \nonumber \\
&&- \mu_0M^0_\mathrm{Co}(1 - p \sin^2\theta){\bf \hat{M}_\mathrm{Co}} \cdot{\bf H }
\label{eq.itin}
\end{eqnarray}
where $K_1$ and $K_2$ quantify the itinerant electron contribution to the anisotropy, and
cos $\theta = {\bf \hat{M}_\mathrm{Co}}\cdot{\bf \hat{z}}$, with  ${\bf \hat{z}}$ pointing along
the $c$ axis.
$p$ quantifies the magnetization anisotropy, which
in the RECo\textsubscript{5} compounds
can cause the Co magnetization to reduce by a few percent 
from its maximum value $M^0_\mathrm{Co}$~\cite{Alameda1981}.
$F_\mathrm{itin}$ depends on temperature through the quantities 
$K_1$, $K_2$, $M^0_\mathrm{Co}$ and $p$.

The two subsystems are coupled together by noting that 
${\bf \hat{M}_\mathrm{Co}} = -{\bf \hat{B}_\mathrm{exch}}$, i.e.\ 
the exchange field felt by the RE-4$f$ electrons points 
antiparallel to the Co magnetization~\cite{Brooks19912}.
The equilibrium direction of ${\bf M_\mathrm{Co}}$ (equivalently,
of ${\bf B_\mathrm{exch}}$) minimizes
the sum of  $F_\mathrm{itin}$ and
$ F_\mathrm{RE}$.
The RE magnetization is obtained as 
${\bf M_\mathrm{RE}} = -\mu_B \langle\hat{\bf L} + 2\hat{\bf S}\rangle_T$,
with $\langle\rangle_T$ denoting the thermal average
over the eigenvalue spectrum of Eq.~\ref{eq.ham} at the equilibrium 
value of ${\bf B_\mathrm{exch}}$, and the total 
magnetization is 
${\bf M_\mathrm{Tot}} = {\bf M_\mathrm{RE}}  + {\bf M_\mathrm{Co}} $.
The magnetization measured along the field direction 
$M$ is ${\bf M_\mathrm{Tot}} \cdot {\bf \hat{H} } $,
whilst the easy direction of magnetization $\alpha$ is obtained
as $\cos^{-1} ( {\bf \hat{M}_\mathrm{Tot}} \cdot {\bf \hat{z}})$
in zero external field. 

${\bf B_\mathrm{exch}}$, $K_1$, $K_2$, $M^0_\mathrm{Co}$ and $p$ 
are obtained
using the FPMVB procedure developed to calculate \emph{F}irst-\emph{P}rinciples
\emph{M}agnetization \emph{V}s \emph{B}-field curves for GdCo\textsubscript{5}~\cite{Patrick2018}.
FPMVB fits DFT-DLM torque calculations~\cite{Staunton2006} 
for different magnetic configurations to extract the desired quantities~\cite{Patrick2018,Patrick20183}.
For all RE\textsuperscript{3+}Co\textsubscript{5} compounds considered
here, we exploit the isovalence of the RE$^{3+}$ ions and substitute the RE 
with Gd in the FPMVB calculations.
This step ensures no erroneous double counting of the CF contribution
which is already accounted for by $F_\mathrm{RE}$,
but still captures the coupling between TM-3$d$ and RE-5$d$ valence states.
We do however use the (experimental) lattice parameters appropriate for each RE~\cite{AndreevHMM,Buschow1977}.
$K_1$, $K_2$, $M^0_\mathrm{Co}$ and $p$ were fitted to
calculations where $\theta$ was varied between 0--90$^\circ$
in 10$^\circ$ intervals.
${\bf B_\mathrm{exch}}$  was obtained by fitting 
the torque induced by introducing a 1$^\circ$ canting between the Gd
and Co sublattices to a free energy of the form $-{\bf B_\mathrm{exch}}\cdot{\bf M_\mathrm{Gd,s}}$
where ${\bf M_\mathrm{Gd,s}}$ is the spin moment of Gd including thermal disorder,
i.e.\ the local spin moment  weighted by the Gd order parameter~\cite{Patrick20183}.
For itinerant CeCo\textsubscript{5} and PrCo\textsubscript{5} (see below), the quantities
in Eq.~\ref{eq.itin} were obtained directly from DFT-DLM calculations on
these compounds.
The calculations used the atomic sphere approximation, angular
momentum expansions with maximum $l$ = 3, and an adaptive reciprocal space
sampling to ensure high precision~\cite{Bruno1997}.
Exchange and correlation were modelled within the local spin-density
approximation (LSDA)~\cite{Vosko1980}, with an orbital
polarization correction applied to the Co-$d$ electrons~\cite{Eriksson19901}
and the LSIC applied to Gd.
The calculated quantities are given as Supplemental Material~\cite{SM}.

We calculate the RECo\textsubscript{5} CF coefficients using an
yttrium-analogue model~\cite{Patrick20192}.
The basic premise here is that due to the isovalence of RE$^{3+}$ ions,
the RE$^{3+}$Co\textsubscript{5} CF potential (which originates from the valence electronic
structure) can be substituted with that of Y$^{3+}$Co\textsubscript{5}.
This step ensures no double-counting of RE-4$f$ electrons, and allows
the use of projector-augmented wave-based DFT calculations to calculate
the CF potential to high accuracy without needing special methods to treat the 4$f$ electrons~\cite{Patrick20192}. 
The CF potential is combined with the radial RE-4$f$ wavefunctions obtained in
LSIC calculations.
At the RE site (symmetry $D_{6h}$) there are four independent components of
the CF potential which affect the 4$f$ anisotropy, 
with $(l,m)$ = (2,0), (4,0), (6,0) and (6,6)[=(6-6)].
The calculated CF coefficients are given as Supplemental
Material~\cite{SM}.
We note that this method implicitly neglects any temperature dependence of 
the CF coefficients themselves, and future work must evaluate
the effects of finite temperature, e.g.\ due to charge fluctuations or lattice expansion~\cite{AndreevHMM}.
The calculations were performed using the 
\texttt{GPAW} code~\cite{Enkovaara2010} within
the LSDA.
A plane wave basis set with a kinetic energy cutoff of 1200~eV
was used, and reciprocal space sampling performed
on a 20$\times$20$\times$20 grid.
The spin-orbit parameter $\lambda$ was calculated using
the RE-centred spherical potential $V_0(r)$ from the LSIC
calculation as $\lambda = \int dr r^2 n^0_{4f}(r) \zeta(r) / (2S)$~\cite{Elliottbook},
where the normalized spherically-symmetric 4$f$ density $n^0_{4f}(r)$
was also obtained from the LSIC calculation~\cite{Patrick20192}
and $\zeta(r) = \frac{\hbar^2}{2m^2c^2}\frac{1}{r}\frac{\partial V_0}{\partial r}$.
The calculated values of $\lambda$ are 1205, 703, 540 and 417~K for Ce, Pr, Nd
and Sm respectively.
These $\lambda$ values yield anisotropy constants indistinguishable
from those calculated using experimental $\lambda$  values
extracted from spectroscopic measurements~\cite{Tiesong1991,Elliottbook}.

\begin{figure}
\includegraphics{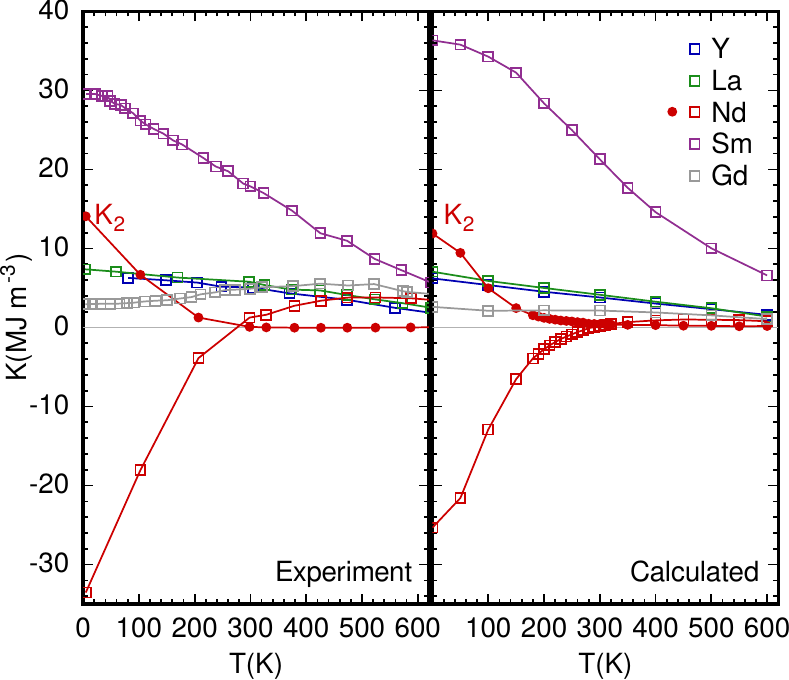}
\caption{Experimental anisotropy constants $\kappa_1$ (and $\kappa_2$ for NdCo\textsubscript{5})~\cite{Ermolenko1976} 
of RECo\textsubscript{5}, compared to the current calculations.
\label{fig.2}}
\end{figure}

We now demonstrate the theory 
by calculating experimentally-measurable quantities.
Figure~\ref{fig.2} (left panel) reproduces anisotropy
constants measured by Ermolenko in 1976~\cite{Ermolenko1976} 
for YCo\textsubscript{5}, 
LaCo\textsubscript{5},
NdCo\textsubscript{5}, 
SmCo\textsubscript{5} and
GdCo\textsubscript{5}.
They were extracted using the Sucksmith-Thompson (ST)
method~\cite{Sucksmith1954},
which is based on the expression for the dependence
of the free energy of a uniaxial ferromagnet on the magnetization
direction $\Theta$:
\begin{equation}
F_\mathrm{FM}(\Theta) = \kappa_1\sin^2\Theta + \kappa_2\sin^4\Theta
+ \mathcal{O}(\sin^6\Theta)
\label{eq.FM}
\end{equation}
As explained in detail in the Supplemental Material~\cite{SM},
measuring the magnetization along the hard direction
and plotting the data as an Arrott plot
($H/M$ vs. $M^2$)~\cite{Arrott1957}
allows $\kappa_1$ and $\kappa_2$ to be extracted
from the gradient and intercept.
Equation~\ref{eq.FM} and the ST method strictly apply
to ferromagnets, but the same technical procedure can be applied 
to RE-TM \emph{ferri}magnets too~\cite{Patrick2018}.
However, the fact that the external field can induce a canting
between the RE and TM moments means that the extracted anisotropy 
constants for the ferrimagnet are effective ones,
which measure both the anisotropy of the individual sublattices
and the strength of the exchange interaction keeping
the spin moments antialigned~\cite{Patrick2018,Patrick20183}.

The experimental data in Fig.~\ref{fig.2} demonstrates the diversity in $\kappa$
among RECo\textsubscript{5}.
The behavior of YCo\textsubscript{5} and LaCo\textsubscript{5},
where the RE is nonmagnetic, is rather similar.
Both compounds display uniaxial anisotropy associated with
the itinerant electron subsystem.
GdCo\textsubscript{5} is still uniaxial, but is softer
than YCo\textsubscript{5} and LaCo\textsubscript{5}.
Since the filled Gd-4$f$ subshell makes no CF
contribution to the anisotropy, this
reduction in $\kappa_1$ is due to the field-induced
canting of the Gd and Co magnetic moments~\cite{Patrick2018}.
SmCo\textsubscript{5} stands out for having the largest
uniaxial anisotropy over the entire temperature range,
with a room temperature value of 17.9~MJm$^{-3}$~\cite{Ermolenko1976}.
NdCo\textsubscript{5} has a negative $\kappa_1$
at low temperatures which switches to positive at 
approximately 280~K,
and also has a non-negligible $\kappa_2$,
at variance with the other compounds.
As discussed below, this variation results
in NdCo\textsubscript{5} undergoing spin reorientation transitions
from planar$\rightarrow$cone$\rightarrow$uniaxial anisotropy~\cite{Klein1975}.

The right panel of Fig.~\ref{fig.2} is the main result of this work,
showing the anisotropy constants obtained
entirely from first principles.
We calculated
hard axis magnetization curves, and then performed the ST
analysis on the data to extract $\kappa_1$ and $\kappa_2$,
in exact correspondence with the experimental procedure~\cite{SM}.
The calculations reproduce all of the experimentally-observed behavior.
SmCo\textsubscript{5} and NdCo\textsubscript{5} have strong
uniaxial and planar anisotropy at zero temperature, respectively.
NdCo\textsubscript{5} has a non-negligible $\kappa_2$ and a value
of $\kappa_1$ which changes sign between 280--290~K.
YCo\textsubscript{5} and LaCo\textsubscript{5} have uniaxial
anisotropy, with LaCo\textsubscript{5} slightly stronger.
GdCo\textsubscript{5} also has uniaxial anisotropy but is softer,
and has a complicated temperature dependence.

Comparing in more detail, we find agreement between experimental
and calculated $\kappa$ values to within a few MJm$^{-3}$ for all 
but the lowest temperatures, where the classical statistical mechanics of
DFT-DLM calculations leads to inaccuracies~\cite{Patrick2017},
and high temperatures, where experimentally the compounds might
undergo decomposition~\cite{Patrick2018}.
We calculate
$\kappa_1$ at room temperature for SmCo\textsubscript{5} to be
21.3~MJm$^{-3}$ (experiment 17.9~MJm$^{-3}$).
Even at zero temperature 
the value of 36.3~MJm$^{-3}$
shows improved agreement with experiment (29.5~MJm$^{-3}$) 
compared to open core (19.7~MJm$^{-3}$)~\cite{Soderlind2017} or  Hubbard+$U$
calculations (40.5~MJm$^{-3}$)~\cite{Larson2004}.

The calculations also reproduce more subtle features,
for instance the slightly enhanced anisotropy (by less than 1~MJm$^{-3}$)
of LaCo\textsubscript{5} over YCo\textsubscript{5}.
The least good agreement is for GdCo\textsubscript{5}, especially
at higher temperatures; however, more recent measurements of $\kappa_1$
found different behavior at elevated temperatures~\cite{Patrick2018,Ermolenko1976}.
At lower temperatures, we note that the present calculations do not include the magnetostatic 
dipole-dipole contribution to the MCA, or the Gd-5$d$ contribution to the itinerant
electron anisotropy, which we previously calculated
to be 24\% the size of the Co contribution~\cite{Patrick2018}.
We conclude that omitting the magnetostatic and RE-$d$ contribution to the 
anisotropy is reasonable for nonmagnetic REs or those with 
unfilled 4$f$ shells (whose RE-4$f$
anisotropy is much larger), but is less suitable for Gd-based
magnets.

\begin{figure}
\includegraphics{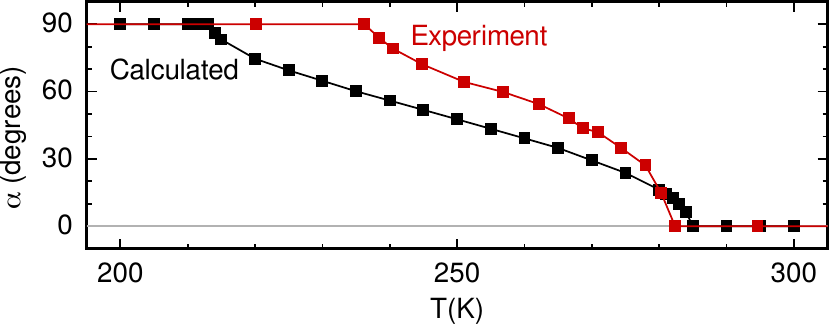}
\caption{
Evolution of the easy magnetization direction $\alpha$ in NdCo\textsubscript{5},
as reported experimentally~\cite{Klein1975} and calculated here.
\label{fig.3}}
\end{figure}
Unlike the other materials in Fig.~\ref{fig.2}, the easy direction of 
magnetization of NdCo\textsubscript{5} lies in the $ab$ plane at low temperature,
with polar angle $\alpha$ = 90$^\circ$.
The anisotropy within the $ab$ plane is determined by the $B_{6\pm6}$ CF coefficients.
Both our calculations and experiments find the easy direction 
to be the $a$ axis, which points from the RE to
between its nearest neighbour Co atoms~\cite{Radwanski1987}.
Experimentally, as $T$ is increased from zero to past approximately 240~K,
a spin reorientation transition (SRT) occurs 
and the magnetization begins to rotate towards the $c$-axis, i.e.\ 
planar$\rightarrow$cone anisotropy.
This rotation continues (decreasing $\alpha$) until approximately
280~K, when a second SRT (cone$\rightarrow$uniaxial) occurs.
Further increasing $T$ sees
$\alpha$ remaining at 0$^\circ$ up to $T_\mathrm{C}$.
The presence of the SRTs close to room temperature led to the
proposal that NdCo\textsubscript{5} may be a candidate material
for magnetic refrigeration~\cite{Nikitin2010}.
The evolution of $\alpha$ as measured experimentally in Ref.~\citenum{Klein1975}
is shown in Fig.~\ref{fig.3}.

The calculated variation of $\alpha$ with temperature is also shown in Fig.~\ref{fig.3}.
We see that the agreement with experiment is remarkably good,
with calculated SRT temperatures of $T_\mathrm{SRT1}$ = 214~K
(plane$\rightarrow$cone) and and $T_\mathrm{SRT2}$ = 285~K (cone$\rightarrow$uniaxial).
There is also good agreement between calculated and experimental $\kappa$
values, especially in the SRT region.
Indeed, the SRTs are intimately linked to the temperature dependence of $\kappa_1$ and $\kappa_2$,
with the plane-cone SRT occurring when $\kappa_1 = -2\kappa_2$ 
and the cone-axis SRT occurring when $\kappa_1$ crosses zero~\cite{Ohkoshi1976,Santosh}.

The calculations provide the underlying physical explanation
of the SRTs, which result from a competition between the
uniaxial anisotropy of the itinerant electrons and a preference
for the oblate Nd$^{3+}$ charge cloud to have its axis
lying in the $ab$ plane (Fig.~\ref{fig.1}).
As the temperature increases, the Nd moments disorder more
quickly than the Co, due to the relative weakness of the
RE-TM exchange field ${\bf B_\mathrm{exch}}$ compared to the
exchange interaction between itinerant moments~\cite{Patrick20182}.
As a result, the negative contribution to $\kappa_1$ from Nd 
reduces quickly with temperature, leaving the positive uniaxial
contribution from the itinerant electrons. 
Obtaining realistic SRT temperatures therefore
requires accounting for the itinerant electron
anisotropy, the crystal field potential and the exchange field
at a comparable level of accuracy.

\begin{figure}
\includegraphics{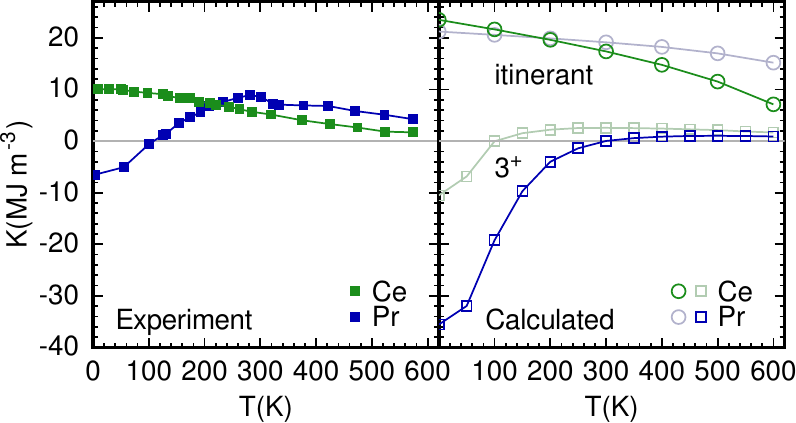}
\caption{Comparison of experimental and calculated anisotropy
constants for CeCo\textsubscript{5} and PrCo\textsubscript{5}.
Values are shown for both itinerant  and
localized (RE in 3+ state) 4$f$ electrons,
with the ground state predicted by the LSIC calculations
shown in the darker color.
\label{fig.4}}
\end{figure}

We finally consider CeCo\textsubscript{5} and PrCo\textsubscript{5}.
Ce has a well-known tendency to undergo transitions between trivalent and tetravalent valence states,
as seen for instance in the $\alpha$-$\gamma$ transition~\cite{Herper2017,Lipp2008}.
The LSIC describes strongly-correlated RE-$4f$ electrons, with
them forming a narrow band several eV below the Fermi level~\cite{Patrick20182}.
Without the LSIC, the 4$f$ electrons are less correlated and more itinerant, appearing
as wider bands close to the Fermi level.
The LSIC finds a lower-energy ground state if the
enhanced correlation offsets the energy penalty associated with the stronger localization~\cite{Lueders2005,Patrick20182}.
Of the RECo\textsubscript{5} compounds, the LSIC predicts a higher energy ground state 
only for CeCo\textsubscript{5}, implying that the Ce-4$f$ electron
is not strongly localized.

Practically, we model compounds with more itinerant (weakly correlated)
RE-4$f$ electrons
by performing non-LSIC DFT-DLM calculations on RECo\textsubscript{5},
with only $F_\mathrm{itin}$ contributing to the free energy.
The values of $\kappa_1$ calculated in this way for CeCo\textsubscript{5}
and PrCo\textsubscript{5} are labelled ``itinerant'' in Fig.~\ref{fig.4}.
We also show $\kappa_1$ values labelled ``3$^+$'',  calculated
for strongly localized RE-4$f$ electrons
using the same method as in Fig.~\ref{fig.2}.
The Ce and Pr moments are held collinear to the Co moments
in the itinerant calculations~\cite{SM}.

Figure~\ref{fig.4} shows a dramatic difference in the anisotropy constants
calculated for the different RE-4$f$ valences.
Pr$^{3+}$Co\textsubscript{5} has an $ab$ plane anisotropy
at low temperature, which is stronger than NdCo\textsubscript{5}.
This behavior is in fact expected; the leading crystal field contribution to the MCA
scales as $J(J-\frac{1}{2})\alpha_J$ ($\alpha_J$ being the Stevens factor),
and this quantity is larger for Pr than Nd~\cite{Kuzmin2008}.
The calculated plane$\rightarrow$cone and cone$\rightarrow$uniaxial SRTs 
occur at 235~K and 297~K
respectively, which are higher temperatures than those calculated for NdCo\textsubscript{5}.
Ce$^{3+}$Co\textsubscript{5} meanwhile is calculated
to have cone anisotropy at zero temperature, with $\alpha$ = 80$^\circ$.
A cone$\rightarrow$axis SRT occurs at 100~K, after which the compound has uniaxial
anisotropy.
The presence of only one SRT shows that Ce$^{3+}$ has a weaker planar anisotropy
than Pr$^{3+}$ or Nd$^{3+}$.
This weaker anisotropy is due to CeCo\textsubscript{5} having
a reduced $B_{20}$ CF coefficient, which correlates with 
a contracted $a$ lattice parameter~\cite{SM}.

If instead the RE-4$f$ electrons are treated as itinerant, 
both CeCo\textsubscript{5} and PrCo\textsubscript{5} are
found to have strong uniaxial anisotropy.
At low temperatures CeCo\textsubscript{5} has the higher
value of $\kappa_1$, (23.5~MJm$^{-3}$ at 0~K), while
above 200~K, $\kappa_1$ of PrCo\textsubscript{5} is larger.
The  $\kappa_1$ values exceed those calculated for YCo\textsubscript{5}
and LaCo\textsubscript{5}, showing that 
delocalizing the RE-4$f$ electrons
boosts the uniaxial anisotropy.

The experimental anisotropy constants from Ref.~\citenum{Ermolenko1976}
are also shown in Fig.~\ref{fig.4}.
The experiments support the picture obtained from the total
energy calculations, that the
Ce-$4f$ and Pr-$4f$ electrons are more 
itinerant/localized (weakly/strongly correlated)
respectively.
CeCo\textsubscript{5} has uniaxial anisotropy across the entire
temperature range.
For PrCo\textsubscript{5}, although $\kappa_1$ is negative at low
temperature, its magnitude is smaller than that measured for 
NdCo\textsubscript{5} (-6.5~MJm$^{-3}$ vs.\ -33.5~MJm$^{-3}$ at 4~K).
As a result, at low temperature the easy magnetization direction of
PrCo\textsubscript{5} does not lie in the $ab$ plane, but rather
in a cone with $\alpha$ = 23$^\circ$~\cite{Tatsumoto1971}.
Increasing the temperature decreases
$\alpha$, and a cone-axis SRT occurs at 105~K~\cite{Tatsumoto1971}.
Therefore, although the Pr ions do favour $ab$-plane anisotropy,
their contribution is weaker than from the Nd ions in 
NdCo\textsubscript{5}, at variance with the CF picture.
Overall, in CeCo\textsubscript{5} (PrCo\textsubscript{5}) experiments find a
smaller uniaxial (planar) contribution from Ce (Pr$^{3+}$).
As a result, the calculated uniaxial anisotropy of CeCo\textsubscript{5}
is larger than found experimentally, while the 
plane$\rightarrow$cone SRT of PrCo\textsubscript{5}
at 235~K
is missing from experiments.

The anomalous behavior of PrCo\textsubscript{5}
in the context of CF theory was
identified in Ref.~\cite{Radwanski19862},
where it was proposed
that in PrCo\textsubscript{5}, Pr assumes a mixed valence
state, e.g.\ Pr$^{3.5+}$, whose properties lie between
Pr and Ce.
The calculations shown in Fig.~\ref{fig.4} support this view,
if we make the reasonable assumption that the anisotropic properties of the mixed
valence state are bounded by those of the strongly localized
and more itinerant (strongly and weakly correlated)
Pr-4$f$ electrons.
In a similar way, the experimentally-observed reduction in CeCo\textsubscript{5}
uniaxial anisotropy compared to the calculations could be explained if the Ce-4$f$ electron 
was more localized (correlated) than predicted by the ``itinerant'' calculations.
From Fig.~\ref{fig.4}, such an electron would be
expected to have a reduced contribution
to the uniaxial anisotropy.
Within this picture, encouraging the itineracy of the Ce-4$f$
electron through e.g.\ chemical pressure,
would boost the uniaxial anisotropy of 
CeCo\textsubscript{5}.

Apart from highlighting the 4$f$-electron physics of Ce and Pr,
our calculations serve as a reminder
of the remarkable properties of SmCo\textsubscript{5}.
As well as its huge zero temperature uniaxial anisotropy,
the large spin moment of Sm strengthens its coupling to the 
exchange field, so that the Sm moments stay ordered up
to higher temperatures.
Furthermore, mixing of the higher-$J$ multiplets also 
boosts the anisotropy~\cite{Buschow1974}.
As a result, as shown in Fig.~\ref{fig.2}, the $\kappa_1$
value of SmCo\textsubscript{5} remains larger than that of
YCo\textsubscript{5} and LaCo\textsubscript{5} 
(where the RE is nonmagnetic),
even at 600~K.
Previously we have shown that the electronic structure
of SmCo\textsubscript{5} close to the Fermi level also gives 
it the highest $T_\mathrm{C}$ of the RECo\textsubscript{5}
compounds~\cite{Patrick20182}.

In summary, we have demonstrated a framework
to calculate the finite-temperature MCA of 
RE-TM magnets. 
Combined with the previously established
DFT-DLM method which provides finite-temperature magnetization
and $T_\mathrm{C}$~\cite{Patrick20182}, we have a full framework
to calculate the intrinsic properties of RE-TM magnets
which requires no experimental
input beyond the crystal structure.
The validation of our method on the RECo\textsubscript{5}
magnet class opens the door to tackling other RE-TM
magnets, like Nd-Fe-B, REFe\textsubscript{12} and Sm\textsubscript{2}Co\textsubscript{17}.
The good performance of the calculations for SmCo\textsubscript{5} will allow us 
to propose strategies to improve this magnet,
e.g.\ through modification of the CF potential
and/or exchange field through TM doping or
application of pressure.
More generally, our work realizes the proposal made
two decades ago in Ref.~\cite{Richter1998},
which suggested that rather than trying to compare 
first-principles CF coefficients to experiment (themselves obtained by fitting), 
the comparison should instead be made for anisotropy constants.

\begin{acknowledgments}
The present work forms part of the PRETAMAG project,
funded by the UK Engineering and Physical Sciences Research
Council, Grant No. EP/M028941/1.
We thank H.\ Akai and M.\ Matsumoto for useful discussions.
\end{acknowledgments}


\begin{thebibliography}{65}%
\makeatletter
\providecommand \@ifxundefined [1]{%
 \@ifx{#1\undefined}
}%
\providecommand \@ifnum [1]{%
 \ifnum #1\expandafter \@firstoftwo
 \else \expandafter \@secondoftwo
 \fi
}%
\providecommand \@ifx [1]{%
 \ifx #1\expandafter \@firstoftwo
 \else \expandafter \@secondoftwo
 \fi
}%
\providecommand \natexlab [1]{#1}%
\providecommand \enquote  [1]{``#1''}%
\providecommand \bibnamefont  [1]{#1}%
\providecommand \bibfnamefont [1]{#1}%
\providecommand \citenamefont [1]{#1}%
\providecommand \href@noop [0]{\@secondoftwo}%
\providecommand \href [0]{\begingroup \@sanitize@url \@href}%
\providecommand \@href[1]{\@@startlink{#1}\@@href}%
\providecommand \@@href[1]{\endgroup#1\@@endlink}%
\providecommand \@sanitize@url [0]{\catcode `\\12\catcode `\$12\catcode
  `\&12\catcode `\#12\catcode `\^12\catcode `\_12\catcode `\%12\relax}%
\providecommand \@@startlink[1]{}%
\providecommand \@@endlink[0]{}%
\providecommand \url  [0]{\begingroup\@sanitize@url \@url }%
\providecommand \@url [1]{\endgroup\@href {#1}{\urlprefix }}%
\providecommand \urlprefix  [0]{URL }%
\providecommand \Eprint [0]{\href }%
\providecommand \doibase [0]{https://doi.org/}%
\providecommand \selectlanguage [0]{\@gobble}%
\providecommand \bibinfo  [0]{\@secondoftwo}%
\providecommand \bibfield  [0]{\@secondoftwo}%
\providecommand \translation [1]{[#1]}%
\providecommand \BibitemOpen [0]{}%
\providecommand \bibitemStop [0]{}%
\providecommand \bibitemNoStop [0]{.\EOS\space}%
\providecommand \EOS [0]{\spacefactor3000\relax}%
\providecommand \BibitemShut  [1]{\csname bibitem#1\endcsname}%
\let\auto@bib@innerbib\@empty
\bibitem [{\citenamefont {Nakamura}(2018)}]{Nakamura2017}%
  \BibitemOpen
  \bibfield  {author} {\bibinfo {author} {\bibfnamefont {H.}~\bibnamefont
  {Nakamura}},\ }\bibfield  {title} {\bibinfo {title} {The current and future
  status of rare earth permanent magnets},\ }\href
  {https://doi.org/10.1016/j.scriptamat.2017.11.010} {\bibfield  {journal}
  {\bibinfo  {journal} {Scr.\ Mater.}\ }\textbf {\bibinfo {volume} {154}},\
  \bibinfo {pages} {273} (\bibinfo {year} {2018})}\BibitemShut {NoStop}%
\bibitem [{\citenamefont {Coey}(2011)}]{Coey2011}%
  \BibitemOpen
  \bibfield  {author} {\bibinfo {author} {\bibfnamefont {J.~M.~D.}\
  \bibnamefont {Coey}},\ }\bibfield  {title} {\bibinfo {title} {Hard {M}agnetic
  {M}aterials: {A} {P}erspective},\ }\href
  {https://doi.org/10.1109/TMAG.2011.2166975} {\bibfield  {journal} {\bibinfo
  {journal} {IEEE Trans.\ Magn.}\ }\textbf {\bibinfo {volume} {47}},\ \bibinfo
  {pages} {4671} (\bibinfo {year} {2011})}\BibitemShut {NoStop}%
\bibitem [{\citenamefont {Gutfleisch}\ \emph {et~al.}(2011)\citenamefont
  {Gutfleisch}, \citenamefont {Willard}, \citenamefont {Br\"uck}, \citenamefont
  {Chen}, \citenamefont {Sankar},\ and\ \citenamefont {Liu}}]{Gutfleisch2011}%
  \BibitemOpen
  \bibfield  {author} {\bibinfo {author} {\bibfnamefont {O.}~\bibnamefont
  {Gutfleisch}}, \bibinfo {author} {\bibfnamefont {M.~A.}\ \bibnamefont
  {Willard}}, \bibinfo {author} {\bibfnamefont {E.}~\bibnamefont {Br\"uck}},
  \bibinfo {author} {\bibfnamefont {C.~H.}\ \bibnamefont {Chen}}, \bibinfo
  {author} {\bibfnamefont {S.~G.}\ \bibnamefont {Sankar}},\ and\ \bibinfo
  {author} {\bibfnamefont {J.~P.}\ \bibnamefont {Liu}},\ }\bibfield  {title}
  {\bibinfo {title} {Magnetic {M}aterials and {D}evices for the 21st {C}entury:
  {S}tronger, {L}ighter, and {M}ore {E}nergy {E}fficient},\ }\href
  {https://doi.org/10.1002/adma.201002180} {\bibfield  {journal} {\bibinfo
  {journal} {Adv.\ Mater.}\ }\textbf {\bibinfo {volume} {23}},\ \bibinfo
  {pages} {821} (\bibinfo {year} {2011})}\BibitemShut {NoStop}%
\bibitem [{\citenamefont {Sagawa}\ \emph {et~al.}(1984)\citenamefont {Sagawa},
  \citenamefont {Fujimura}, \citenamefont {Togawa}, \citenamefont {Yamamoto},\
  and\ \citenamefont {Matsuura}}]{Sagawa1984}%
  \BibitemOpen
  \bibfield  {author} {\bibinfo {author} {\bibfnamefont {M.}~\bibnamefont
  {Sagawa}}, \bibinfo {author} {\bibfnamefont {S.}~\bibnamefont {Fujimura}},
  \bibinfo {author} {\bibfnamefont {N.}~\bibnamefont {Togawa}}, \bibinfo
  {author} {\bibfnamefont {H.}~\bibnamefont {Yamamoto}},\ and\ \bibinfo
  {author} {\bibfnamefont {Y.}~\bibnamefont {Matsuura}},\ }\bibfield  {title}
  {\bibinfo {title} {New material for permanent magnets on a base of {N}d and
  {F}e (invited)},\ }\href {https://doi.org/10.1063/1.333572} {\bibfield
  {journal} {\bibinfo  {journal} {J.\ Appl.\ Phys.}\ }\textbf {\bibinfo
  {volume} {55}},\ \bibinfo {pages} {2083} (\bibinfo {year}
  {1984})}\BibitemShut {NoStop}%
\bibitem [{\citenamefont {Croat}\ \emph {et~al.}(1984)\citenamefont {Croat},
  \citenamefont {Herbst}, \citenamefont {Lee},\ and\ \citenamefont
  {Pinkerton}}]{Croat1984}%
  \BibitemOpen
  \bibfield  {author} {\bibinfo {author} {\bibfnamefont {J.~J.}\ \bibnamefont
  {Croat}}, \bibinfo {author} {\bibfnamefont {J.~F.}\ \bibnamefont {Herbst}},
  \bibinfo {author} {\bibfnamefont {R.~W.}\ \bibnamefont {Lee}},\ and\ \bibinfo
  {author} {\bibfnamefont {F.~E.}\ \bibnamefont {Pinkerton}},\ }\bibfield
  {title} {\bibinfo {title} {Pr-{F}e and {N}d-{F}e-based materials: A new class
  of high-performance permanent magnets (invited)},\ }\href
  {https://doi.org/10.1063/1.333571} {\bibfield  {journal} {\bibinfo  {journal}
  {J.\ Appl.\ Phys.}\ }\textbf {\bibinfo {volume} {55}},\ \bibinfo {pages}
  {2078} (\bibinfo {year} {1984})}\BibitemShut {NoStop}%
\bibitem [{\citenamefont {Strnat}(1972)}]{Strnat1972}%
  \BibitemOpen
  \bibfield  {author} {\bibinfo {author} {\bibfnamefont {K.}~\bibnamefont
  {Strnat}},\ }\bibfield  {title} {\bibinfo {title} {The hard-magnetic
  properties of rare earth-transition metal alloys},\ }\href
  {https://doi.org/10.1109/TMAG.1972.1067368} {\bibfield  {journal} {\bibinfo
  {journal} {IEEE Trans.\ Magn.}\ }\textbf {\bibinfo {volume} {8}},\ \bibinfo
  {pages} {511} (\bibinfo {year} {1972})}\BibitemShut {NoStop}%
\bibitem [{\citenamefont {Strnat}\ \emph {et~al.}(1967)\citenamefont {Strnat},
  \citenamefont {Hoffer}, \citenamefont {Olson}, \citenamefont {Ostertag},\
  and\ \citenamefont {Becker}}]{Strnat1967}%
  \BibitemOpen
  \bibfield  {author} {\bibinfo {author} {\bibfnamefont {K.}~\bibnamefont
  {Strnat}}, \bibinfo {author} {\bibfnamefont {G.}~\bibnamefont {Hoffer}},
  \bibinfo {author} {\bibfnamefont {J.}~\bibnamefont {Olson}}, \bibinfo
  {author} {\bibfnamefont {W.}~\bibnamefont {Ostertag}},\ and\ \bibinfo
  {author} {\bibfnamefont {J.~J.}\ \bibnamefont {Becker}},\ }\bibfield  {title}
  {\bibinfo {title} {A family of new cobalt‐base permanent magnet
  materials},\ }\href {https://doi.org/10.1063/1.1709459} {\bibfield  {journal}
  {\bibinfo  {journal} {J. Appl. Phys.}\ }\textbf {\bibinfo {volume} {38}},\
  \bibinfo {pages} {1001} (\bibinfo {year} {1967})}\BibitemShut {NoStop}%
\bibitem [{\citenamefont {Pathak}\ \emph {et~al.}(2015)\citenamefont {Pathak},
  \citenamefont {Khan}, \citenamefont {Gschneidner~Jr.}, \citenamefont
  {McCallum}, \citenamefont {Zhou}, \citenamefont {Sun}, \citenamefont
  {Dennis}, \citenamefont {Zhou}, \citenamefont {Pinkerton}, \citenamefont
  {Kramer},\ and\ \citenamefont {Pecharsky}}]{Pathak2015}%
  \BibitemOpen
  \bibfield  {author} {\bibinfo {author} {\bibfnamefont {A.~K.}\ \bibnamefont
  {Pathak}}, \bibinfo {author} {\bibfnamefont {M.}~\bibnamefont {Khan}},
  \bibinfo {author} {\bibfnamefont {K.~A.}\ \bibnamefont {Gschneidner~Jr.}},
  \bibinfo {author} {\bibfnamefont {R.~W.}\ \bibnamefont {McCallum}}, \bibinfo
  {author} {\bibfnamefont {L.}~\bibnamefont {Zhou}}, \bibinfo {author}
  {\bibfnamefont {K.}~\bibnamefont {Sun}}, \bibinfo {author} {\bibfnamefont
  {K.~W.}\ \bibnamefont {Dennis}}, \bibinfo {author} {\bibfnamefont
  {C.}~\bibnamefont {Zhou}}, \bibinfo {author} {\bibfnamefont {F.~E.}\
  \bibnamefont {Pinkerton}}, \bibinfo {author} {\bibfnamefont {M.~J.}\
  \bibnamefont {Kramer}},\ and\ \bibinfo {author} {\bibfnamefont {V.~K.}\
  \bibnamefont {Pecharsky}},\ }\bibfield  {title} {\bibinfo {title} {Cerium:
  {A}n {U}nlikely {R}eplacement of {D}ysprosium in {H}igh {P}erformance
  {N}d–{F}e–{B} {P}ermanent {M}agnets},\ }\href
  {https://doi.org/10.1002/adma.201404892} {\bibfield  {journal} {\bibinfo
  {journal} {Adv.\ Mater.}\ }\textbf {\bibinfo {volume} {27}},\ \bibinfo
  {pages} {2663} (\bibinfo {year} {2015})}\BibitemShut {NoStop}%
\bibitem [{\citenamefont {Vekilova}\ \emph {et~al.}(2019)\citenamefont
  {Vekilova}, \citenamefont {Fayyazi}, \citenamefont {Skokov}, \citenamefont
  {Gutfleisch}, \citenamefont {Echevarria-Bonet}, \citenamefont
  {Barandiar\'an}, \citenamefont {Kovacs}, \citenamefont {Fischbacher},
  \citenamefont {Schrefl}, \citenamefont {Eriksson},\ and\ \citenamefont
  {Herper}}]{Vekilova2019}%
  \BibitemOpen
  \bibfield  {author} {\bibinfo {author} {\bibfnamefont {O.~Y.}\ \bibnamefont
  {Vekilova}}, \bibinfo {author} {\bibfnamefont {B.}~\bibnamefont {Fayyazi}},
  \bibinfo {author} {\bibfnamefont {K.~P.}\ \bibnamefont {Skokov}}, \bibinfo
  {author} {\bibfnamefont {O.}~\bibnamefont {Gutfleisch}}, \bibinfo {author}
  {\bibfnamefont {C.}~\bibnamefont {Echevarria-Bonet}}, \bibinfo {author}
  {\bibfnamefont {J.~M.}\ \bibnamefont {Barandiar\'an}}, \bibinfo {author}
  {\bibfnamefont {A.}~\bibnamefont {Kovacs}}, \bibinfo {author} {\bibfnamefont
  {J.}~\bibnamefont {Fischbacher}}, \bibinfo {author} {\bibfnamefont
  {T.}~\bibnamefont {Schrefl}}, \bibinfo {author} {\bibfnamefont
  {O.}~\bibnamefont {Eriksson}},\ and\ \bibinfo {author} {\bibfnamefont
  {H.~C.}\ \bibnamefont {Herper}},\ }\bibfield  {title} {\bibinfo {title}
  {Tuning the magnetocrystalline anisotropy of {F}e$_3${S}n by alloying},\
  }\href {https://doi.org/10.1103/PhysRevB.99.024421} {\bibfield  {journal}
  {\bibinfo  {journal} {Phys. Rev. B}\ }\textbf {\bibinfo {volume} {99}},\
  \bibinfo {pages} {024421} (\bibinfo {year} {2019})}\BibitemShut {NoStop}%
\bibitem [{\citenamefont {Tatetsu}\ \emph {et~al.}(2018)\citenamefont
  {Tatetsu}, \citenamefont {Harashima}, \citenamefont {Miyake},\ and\
  \citenamefont {Gohda}}]{Tatetsu2018}%
  \BibitemOpen
  \bibfield  {author} {\bibinfo {author} {\bibfnamefont {Y.}~\bibnamefont
  {Tatetsu}}, \bibinfo {author} {\bibfnamefont {Y.}~\bibnamefont {Harashima}},
  \bibinfo {author} {\bibfnamefont {T.}~\bibnamefont {Miyake}},\ and\ \bibinfo
  {author} {\bibfnamefont {Y.}~\bibnamefont {Gohda}},\ }\bibfield  {title}
  {\bibinfo {title} {Role of typical elements in {N}d$_2${F}e$_{14}{X}$
  (${X}$={B},{C},{N},{O},{F})},\ }\href
  {https://doi.org/10.1103/PhysRevMaterials.2.074410} {\bibfield  {journal}
  {\bibinfo  {journal} {Phys. Rev. Mater.}\ }\textbf {\bibinfo {volume} {2}},\
  \bibinfo {pages} {074410} (\bibinfo {year} {2018})}\BibitemShut {NoStop}%
\bibitem [{\citenamefont {Perdew}\ and\ \citenamefont
  {Zunger}(1981)}]{Perdew1981}%
  \BibitemOpen
  \bibfield  {author} {\bibinfo {author} {\bibfnamefont {J.~P.}\ \bibnamefont
  {Perdew}}\ and\ \bibinfo {author} {\bibfnamefont {A.}~\bibnamefont
  {Zunger}},\ }\bibfield  {title} {\bibinfo {title} {Self-interaction
  correction to density-functional approximations for many-electron systems},\
  }\href {https://doi.org/10.1103/PhysRevB.23.5048} {\bibfield  {journal}
  {\bibinfo  {journal} {Phys. Rev. B}\ }\textbf {\bibinfo {volume} {23}},\
  \bibinfo {pages} {5048} (\bibinfo {year} {1981})}\BibitemShut {NoStop}%
\bibitem [{\citenamefont {Kohn}\ and\ \citenamefont {Sham}(1965)}]{Kohn1965}%
  \BibitemOpen
  \bibfield  {author} {\bibinfo {author} {\bibfnamefont {W.}~\bibnamefont
  {Kohn}}\ and\ \bibinfo {author} {\bibfnamefont {L.~J.}\ \bibnamefont
  {Sham}},\ }\bibfield  {title} {\bibinfo {title} {Self-consistent equations
  including exchange and correlation effects},\ }\href
  {https://doi.org/10.1103/PhysRev.140.A1133} {\bibfield  {journal} {\bibinfo
  {journal} {Phys. Rev.}\ }\textbf {\bibinfo {volume} {140}},\ \bibinfo {pages}
  {A1133} (\bibinfo {year} {1965})}\BibitemShut {NoStop}%
\bibitem [{\citenamefont {Locht}\ \emph {et~al.}(2016)\citenamefont {Locht},
  \citenamefont {Kvashnin}, \citenamefont {Rodrigues}, \citenamefont {Pereiro},
  \citenamefont {Bergman}, \citenamefont {Bergqvist}, \citenamefont
  {Lichtenstein}, \citenamefont {Katsnelson}, \citenamefont {Delin},
  \citenamefont {Klautau}, \citenamefont {Johansson}, \citenamefont
  {Di~Marco},\ and\ \citenamefont {Eriksson}}]{Locht2016}%
  \BibitemOpen
  \bibfield  {author} {\bibinfo {author} {\bibfnamefont {I.~L.~M.}\
  \bibnamefont {Locht}}, \bibinfo {author} {\bibfnamefont {Y.~O.}\ \bibnamefont
  {Kvashnin}}, \bibinfo {author} {\bibfnamefont {D.~C.~M.}\ \bibnamefont
  {Rodrigues}}, \bibinfo {author} {\bibfnamefont {M.}~\bibnamefont {Pereiro}},
  \bibinfo {author} {\bibfnamefont {A.}~\bibnamefont {Bergman}}, \bibinfo
  {author} {\bibfnamefont {L.}~\bibnamefont {Bergqvist}}, \bibinfo {author}
  {\bibfnamefont {A.~I.}\ \bibnamefont {Lichtenstein}}, \bibinfo {author}
  {\bibfnamefont {M.~I.}\ \bibnamefont {Katsnelson}}, \bibinfo {author}
  {\bibfnamefont {A.}~\bibnamefont {Delin}}, \bibinfo {author} {\bibfnamefont
  {A.~B.}\ \bibnamefont {Klautau}}, \bibinfo {author} {\bibfnamefont
  {B.}~\bibnamefont {Johansson}}, \bibinfo {author} {\bibfnamefont
  {I.}~\bibnamefont {Di~Marco}},\ and\ \bibinfo {author} {\bibfnamefont
  {O.}~\bibnamefont {Eriksson}},\ }\bibfield  {title} {\bibinfo {title}
  {Standard model of the rare earths analyzed from the {H}ubbard {I}
  approximation},\ }\href {https://doi.org/10.1103/PhysRevB.94.085137}
  {\bibfield  {journal} {\bibinfo  {journal} {Phys. Rev. B}\ }\textbf {\bibinfo
  {volume} {94}},\ \bibinfo {pages} {085137} (\bibinfo {year}
  {2016})}\BibitemShut {NoStop}%
\bibitem [{\citenamefont {Delange}\ \emph {et~al.}(2017)\citenamefont
  {Delange}, \citenamefont {Biermann}, \citenamefont {Miyake},\ and\
  \citenamefont {Pourovskii}}]{Delange2017}%
  \BibitemOpen
  \bibfield  {author} {\bibinfo {author} {\bibfnamefont {P.}~\bibnamefont
  {Delange}}, \bibinfo {author} {\bibfnamefont {S.}~\bibnamefont {Biermann}},
  \bibinfo {author} {\bibfnamefont {T.}~\bibnamefont {Miyake}},\ and\ \bibinfo
  {author} {\bibfnamefont {L.}~\bibnamefont {Pourovskii}},\ }\bibfield  {title}
  {\bibinfo {title} {Crystal-field splittings in rare-earth-based hard magnets:
  An ab initio approach},\ }\href {https://doi.org/10.1103/PhysRevB.96.155132}
  {\bibfield  {journal} {\bibinfo  {journal} {Phys. Rev. B}\ }\textbf {\bibinfo
  {volume} {96}},\ \bibinfo {pages} {155132} (\bibinfo {year}
  {2017})}\BibitemShut {NoStop}%
\bibitem [{\citenamefont {L\"uders}\ \emph {et~al.}(2005)\citenamefont
  {L\"uders}, \citenamefont {Ernst}, \citenamefont {D\"ane}, \citenamefont
  {Szotek}, \citenamefont {Svane}, \citenamefont {K\"odderitzsch},
  \citenamefont {Hergert}, \citenamefont {Gy\"orffy},\ and\ \citenamefont
  {Temmerman}}]{Lueders2005}%
  \BibitemOpen
  \bibfield  {author} {\bibinfo {author} {\bibfnamefont {M.}~\bibnamefont
  {L\"uders}}, \bibinfo {author} {\bibfnamefont {A.}~\bibnamefont {Ernst}},
  \bibinfo {author} {\bibfnamefont {M.}~\bibnamefont {D\"ane}}, \bibinfo
  {author} {\bibfnamefont {Z.}~\bibnamefont {Szotek}}, \bibinfo {author}
  {\bibfnamefont {A.}~\bibnamefont {Svane}}, \bibinfo {author} {\bibfnamefont
  {D.}~\bibnamefont {K\"odderitzsch}}, \bibinfo {author} {\bibfnamefont
  {W.}~\bibnamefont {Hergert}}, \bibinfo {author} {\bibfnamefont {B.~L.}\
  \bibnamefont {Gy\"orffy}},\ and\ \bibinfo {author} {\bibfnamefont {W.~M.}\
  \bibnamefont {Temmerman}},\ }\bibfield  {title} {\bibinfo {title}
  {Self-interaction correction in multiple scattering theory},\ }\href
  {https://doi.org/10.1103/PhysRevB.71.205109} {\bibfield  {journal} {\bibinfo
  {journal} {Phys. Rev. B}\ }\textbf {\bibinfo {volume} {71}},\ \bibinfo
  {pages} {205109} (\bibinfo {year} {2005})}\BibitemShut {NoStop}%
\bibitem [{\citenamefont {Brooks}\ \emph {et~al.}(1991)\citenamefont {Brooks},
  \citenamefont {Nordstrom},\ and\ \citenamefont {Johansson}}]{Brooks19912}%
  \BibitemOpen
  \bibfield  {author} {\bibinfo {author} {\bibfnamefont {M.~S.~S.}\
  \bibnamefont {Brooks}}, \bibinfo {author} {\bibfnamefont {L.}~\bibnamefont
  {Nordstrom}},\ and\ \bibinfo {author} {\bibfnamefont {B.}~\bibnamefont
  {Johansson}},\ }\bibfield  {title} {\bibinfo {title} {3$d$-5$d$ band
  magnetism in rare earth-transition metal intermetallics: total and partial
  magnetic moments of the {RF}e\textsubscript{2} ({R}={G}d-{Y}b) {L}aves phase
  compounds},\ }\href@noop {} {\bibfield  {journal} {\bibinfo  {journal} {J.
  Phys.: Condens. Matter}\ }\textbf {\bibinfo {volume} {3}},\ \bibinfo {pages}
  {2357} (\bibinfo {year} {1991})}\BibitemShut {NoStop}%
\bibitem [{\citenamefont {S\"oderlind}\ \emph {et~al.}(2017)\citenamefont
  {S\"oderlind}, \citenamefont {Landa}, \citenamefont {Locht}, \citenamefont
  {\AA{}berg}, \citenamefont {Kvashnin}, \citenamefont {Pereiro}, \citenamefont
  {D\"ane}, \citenamefont {Turchi}, \citenamefont {Antropov},\ and\
  \citenamefont {Eriksson}}]{Soderlind2017}%
  \BibitemOpen
  \bibfield  {author} {\bibinfo {author} {\bibfnamefont {P.}~\bibnamefont
  {S\"oderlind}}, \bibinfo {author} {\bibfnamefont {A.}~\bibnamefont {Landa}},
  \bibinfo {author} {\bibfnamefont {I.~L.~M.}\ \bibnamefont {Locht}}, \bibinfo
  {author} {\bibfnamefont {D.}~\bibnamefont {\AA{}berg}}, \bibinfo {author}
  {\bibfnamefont {Y.}~\bibnamefont {Kvashnin}}, \bibinfo {author}
  {\bibfnamefont {M.}~\bibnamefont {Pereiro}}, \bibinfo {author} {\bibfnamefont
  {M.}~\bibnamefont {D\"ane}}, \bibinfo {author} {\bibfnamefont {P.~E.~A.}\
  \bibnamefont {Turchi}}, \bibinfo {author} {\bibfnamefont {V.~P.}\
  \bibnamefont {Antropov}},\ and\ \bibinfo {author} {\bibfnamefont
  {O.}~\bibnamefont {Eriksson}},\ }\bibfield  {title} {\bibinfo {title}
  {Prediction of the new efficient permanent magnet
  {S}m{C}o{N}i{F}e\textsubscript{3}},\ }\href
  {https://doi.org/10.1103/PhysRevB.96.100404} {\bibfield  {journal} {\bibinfo
  {journal} {Phys. Rev. B}\ }\textbf {\bibinfo {volume} {96}},\ \bibinfo
  {pages} {100404} (\bibinfo {year} {2017})}\BibitemShut {NoStop}%
\bibitem [{\citenamefont {Fukazawa}\ \emph {et~al.}(2017)\citenamefont
  {Fukazawa}, \citenamefont {Akai}, \citenamefont {Harashima},\ and\
  \citenamefont {Miyake}}]{Fukazawa2017}%
  \BibitemOpen
  \bibfield  {author} {\bibinfo {author} {\bibfnamefont {T.}~\bibnamefont
  {Fukazawa}}, \bibinfo {author} {\bibfnamefont {H.}~\bibnamefont {Akai}},
  \bibinfo {author} {\bibfnamefont {Y.}~\bibnamefont {Harashima}},\ and\
  \bibinfo {author} {\bibfnamefont {T.}~\bibnamefont {Miyake}},\ }\bibfield
  {title} {\bibinfo {title} {First-principles study of intersite magnetic
  couplings in {N}d{F}e\textsubscript{12} and {N}d{F}e\textsubscript{12}{X}
  ({X} = {B, C, N, O, F})},\ }\href {https://doi.org/10.1063/1.4996989}
  {\bibfield  {journal} {\bibinfo  {journal} {J. Appl. Phys.}\ }\textbf
  {\bibinfo {volume} {122}},\ \bibinfo {pages} {053901} (\bibinfo {year}
  {2017})}\BibitemShut {NoStop}%
\bibitem [{\citenamefont {Waller}\ \emph {et~al.}(2016)\citenamefont {Waller},
  \citenamefont {Piekarz}, \citenamefont {Bosak}, \citenamefont {Jochym},
  \citenamefont {Ibrahimkutty}, \citenamefont {Seiler}, \citenamefont {Krisch},
  \citenamefont {Baumbach}, \citenamefont {Parlinski},\ and\ \citenamefont
  {Stankov}}]{Waller2016}%
  \BibitemOpen
  \bibfield  {author} {\bibinfo {author} {\bibfnamefont {O.}~\bibnamefont
  {Waller}}, \bibinfo {author} {\bibfnamefont {P.}~\bibnamefont {Piekarz}},
  \bibinfo {author} {\bibfnamefont {A.}~\bibnamefont {Bosak}}, \bibinfo
  {author} {\bibfnamefont {P.~T.}\ \bibnamefont {Jochym}}, \bibinfo {author}
  {\bibfnamefont {S.}~\bibnamefont {Ibrahimkutty}}, \bibinfo {author}
  {\bibfnamefont {A.}~\bibnamefont {Seiler}}, \bibinfo {author} {\bibfnamefont
  {M.}~\bibnamefont {Krisch}}, \bibinfo {author} {\bibfnamefont
  {T.}~\bibnamefont {Baumbach}}, \bibinfo {author} {\bibfnamefont
  {K.}~\bibnamefont {Parlinski}},\ and\ \bibinfo {author} {\bibfnamefont
  {S.}~\bibnamefont {Stankov}},\ }\bibfield  {title} {\bibinfo {title} {Lattice
  dynamics of neodymium: Influence of $4f$ electron correlations},\ }\href
  {https://doi.org/10.1103/PhysRevB.94.014303} {\bibfield  {journal} {\bibinfo
  {journal} {Phys. Rev. B}\ }\textbf {\bibinfo {volume} {94}},\ \bibinfo
  {pages} {014303} (\bibinfo {year} {2016})}\BibitemShut {NoStop}%
\bibitem [{\citenamefont {Larson}\ \emph {et~al.}(2004)\citenamefont {Larson},
  \citenamefont {Mazin},\ and\ \citenamefont
  {Papaconstantopoulos}}]{Larson2004}%
  \BibitemOpen
  \bibfield  {author} {\bibinfo {author} {\bibfnamefont {P.}~\bibnamefont
  {Larson}}, \bibinfo {author} {\bibfnamefont {I.~I.}\ \bibnamefont {Mazin}},\
  and\ \bibinfo {author} {\bibfnamefont {D.~A.}\ \bibnamefont
  {Papaconstantopoulos}},\ }\bibfield  {title} {\bibinfo {title} {Effects of
  doping on the magnetic anisotropy energy in
  {SmCo}\textsubscript{5-x}{Fe}\textsubscript{x} and
  {YCo}\textsubscript{5-x}{Fe}\textsubscript{x}},\ }\href
  {https://doi.org/10.1103/PhysRevB.69.134408} {\bibfield  {journal} {\bibinfo
  {journal} {Phys. Rev. B}\ }\textbf {\bibinfo {volume} {69}},\ \bibinfo
  {pages} {134408} (\bibinfo {year} {2004})}\BibitemShut {NoStop}%
\bibitem [{\citenamefont {Gy\"orffy}\ \emph {et~al.}(1985)\citenamefont
  {Gy\"orffy}, \citenamefont {Pindor}, \citenamefont {Staunton}, \citenamefont
  {Stocks},\ and\ \citenamefont {Winter}}]{Gyorffy1985}%
  \BibitemOpen
  \bibfield  {author} {\bibinfo {author} {\bibfnamefont {B.~L.}\ \bibnamefont
  {Gy\"orffy}}, \bibinfo {author} {\bibfnamefont {A.~J.}\ \bibnamefont
  {Pindor}}, \bibinfo {author} {\bibfnamefont {J.}~\bibnamefont {Staunton}},
  \bibinfo {author} {\bibfnamefont {G.~M.}\ \bibnamefont {Stocks}},\ and\
  \bibinfo {author} {\bibfnamefont {H.}~\bibnamefont {Winter}},\ }\bibfield
  {title} {\bibinfo {title} {A first-principles theory of ferromagnetic phase
  transitions in metals},\ }\href {https://doi.org/10.1088/0305-4608/15/6/018}
  {\bibfield  {journal} {\bibinfo  {journal} {J. Phys. F: Met. Phys.}\ }\textbf
  {\bibinfo {volume} {15}},\ \bibinfo {pages} {1337} (\bibinfo {year}
  {1985})}\BibitemShut {NoStop}%
\bibitem [{\citenamefont {Hughes}\ \emph {et~al.}(2007)\citenamefont {Hughes},
  \citenamefont {Daane}, \citenamefont {Ernst}, \citenamefont {Hergert},
  \citenamefont {L\"uders}, \citenamefont {Poulter}, \citenamefont {Staunton},
  \citenamefont {Svane}, \citenamefont {Szotek},\ and\ \citenamefont
  {Temmerman}}]{Hughes2007}%
  \BibitemOpen
  \bibfield  {author} {\bibinfo {author} {\bibfnamefont {I.~D.}\ \bibnamefont
  {Hughes}}, \bibinfo {author} {\bibfnamefont {M.}~\bibnamefont {Daane}},
  \bibinfo {author} {\bibfnamefont {A.}~\bibnamefont {Ernst}}, \bibinfo
  {author} {\bibfnamefont {W.}~\bibnamefont {Hergert}}, \bibinfo {author}
  {\bibfnamefont {M.}~\bibnamefont {L\"uders}}, \bibinfo {author}
  {\bibfnamefont {J.}~\bibnamefont {Poulter}}, \bibinfo {author} {\bibfnamefont
  {J.~B.}\ \bibnamefont {Staunton}}, \bibinfo {author} {\bibfnamefont
  {A.}~\bibnamefont {Svane}}, \bibinfo {author} {\bibfnamefont
  {Z.}~\bibnamefont {Szotek}},\ and\ \bibinfo {author} {\bibfnamefont {W.~M.}\
  \bibnamefont {Temmerman}},\ }\bibfield  {title} {\bibinfo {title} {Lanthanide
  contraction and magnetism in the heavy rare earth elements},\ }\href
  {https://doi.org/10.1038/nature05668} {\bibfield  {journal} {\bibinfo
  {journal} {Nature}\ }\textbf {\bibinfo {volume} {446}},\ \bibinfo {pages}
  {650} (\bibinfo {year} {2007})}\BibitemShut {NoStop}%
\bibitem [{\citenamefont {Mendive-Tapia}\ and\ \citenamefont
  {Staunton}(2017)}]{MendiveTapia2017}%
  \BibitemOpen
  \bibfield  {author} {\bibinfo {author} {\bibfnamefont {E.}~\bibnamefont
  {Mendive-Tapia}}\ and\ \bibinfo {author} {\bibfnamefont {J.~B.}\ \bibnamefont
  {Staunton}},\ }\bibfield  {title} {\bibinfo {title} {Theory of magnetic
  ordering in the heavy rare earths: Ab initio electronic origin of pair- and
  four-spin interactions},\ }\href
  {https://doi.org/10.1103/PhysRevLett.118.197202} {\bibfield  {journal}
  {\bibinfo  {journal} {Phys. Rev. Lett.}\ }\textbf {\bibinfo {volume} {118}},\
  \bibinfo {pages} {197202} (\bibinfo {year} {2017})}\BibitemShut {NoStop}%
\bibitem [{\citenamefont {Patrick}\ and\ \citenamefont
  {Staunton}(2018)}]{Patrick20182}%
  \BibitemOpen
  \bibfield  {author} {\bibinfo {author} {\bibfnamefont {C.~E.}\ \bibnamefont
  {Patrick}}\ and\ \bibinfo {author} {\bibfnamefont {J.~B.}\ \bibnamefont
  {Staunton}},\ }\bibfield  {title} {\bibinfo {title}
  {Rare-earth/transition-metal magnets at finite temperature:
  {S}elf-interaction-corrected relativistic density functional theory in the
  disordered local moment picture},\ }\href
  {https://doi.org/10.1103/PhysRevB.97.224415} {\bibfield  {journal} {\bibinfo
  {journal} {Phys. Rev. B}\ }\textbf {\bibinfo {volume} {97}},\ \bibinfo
  {pages} {224415} (\bibinfo {year} {2018})}\BibitemShut {NoStop}%
\bibitem [{\citenamefont {Staunton}\ \emph {et~al.}(2004)\citenamefont
  {Staunton}, \citenamefont {Ostanin}, \citenamefont {Razee}, \citenamefont
  {Gy\"orffy}, \citenamefont {Szunyogh}, \citenamefont {Ginatempo},\ and\
  \citenamefont {Bruno}}]{Staunton2004}%
  \BibitemOpen
  \bibfield  {author} {\bibinfo {author} {\bibfnamefont {J.~B.}\ \bibnamefont
  {Staunton}}, \bibinfo {author} {\bibfnamefont {S.}~\bibnamefont {Ostanin}},
  \bibinfo {author} {\bibfnamefont {S.~S.~A.}\ \bibnamefont {Razee}}, \bibinfo
  {author} {\bibfnamefont {B.~L.}\ \bibnamefont {Gy\"orffy}}, \bibinfo {author}
  {\bibfnamefont {L.}~\bibnamefont {Szunyogh}}, \bibinfo {author}
  {\bibfnamefont {B.}~\bibnamefont {Ginatempo}},\ and\ \bibinfo {author}
  {\bibfnamefont {E.}~\bibnamefont {Bruno}},\ }\bibfield  {title} {\bibinfo
  {title} {Temperature dependent magnetic anisotropy in metallic magnets from
  an \textit{Ab Initio} electronic structure theory: ${L}{1}_{0}$-ordered
  {FePt}},\ }\href {https://doi.org/10.1103/PhysRevLett.93.257204} {\bibfield
  {journal} {\bibinfo  {journal} {Phys. Rev. Lett.}\ }\textbf {\bibinfo
  {volume} {93}},\ \bibinfo {pages} {257204} (\bibinfo {year}
  {2004})}\BibitemShut {NoStop}%
\bibitem [{\citenamefont {Matsumoto}\ \emph {et~al.}(2014)\citenamefont
  {Matsumoto}, \citenamefont {Banerjee},\ and\ \citenamefont
  {Staunton}}]{Matsumoto2014}%
  \BibitemOpen
  \bibfield  {author} {\bibinfo {author} {\bibfnamefont {M.}~\bibnamefont
  {Matsumoto}}, \bibinfo {author} {\bibfnamefont {R.}~\bibnamefont
  {Banerjee}},\ and\ \bibinfo {author} {\bibfnamefont {J.~B.}\ \bibnamefont
  {Staunton}},\ }\bibfield  {title} {\bibinfo {title} {Improvement of magnetic
  hardness at finite temperatures: \textit{{A}b initio} disordered local-moment
  approach for {YC}o\textsubscript{5}},\ }\href
  {https://doi.org/10.1103/PhysRevB.90.054421} {\bibfield  {journal} {\bibinfo
  {journal} {Phys. Rev. B}\ }\textbf {\bibinfo {volume} {90}},\ \bibinfo
  {pages} {054421} (\bibinfo {year} {2014})}\BibitemShut {NoStop}%
\bibitem [{\citenamefont {Patrick}\ \emph
  {et~al.}(2018{\natexlab{a}})\citenamefont {Patrick}, \citenamefont {Kumar},
  \citenamefont {Balakrishnan}, \citenamefont {Edwards}, \citenamefont {Lees},
  \citenamefont {Petit},\ and\ \citenamefont {Staunton}}]{Patrick2018}%
  \BibitemOpen
  \bibfield  {author} {\bibinfo {author} {\bibfnamefont {C.~E.}\ \bibnamefont
  {Patrick}}, \bibinfo {author} {\bibfnamefont {S.}~\bibnamefont {Kumar}},
  \bibinfo {author} {\bibfnamefont {G.}~\bibnamefont {Balakrishnan}}, \bibinfo
  {author} {\bibfnamefont {R.~S.}\ \bibnamefont {Edwards}}, \bibinfo {author}
  {\bibfnamefont {M.~R.}\ \bibnamefont {Lees}}, \bibinfo {author}
  {\bibfnamefont {L.}~\bibnamefont {Petit}},\ and\ \bibinfo {author}
  {\bibfnamefont {J.~B.}\ \bibnamefont {Staunton}},\ }\bibfield  {title}
  {\bibinfo {title} {Calculating the magnetic anisotropy of
  rare-earth---transition-metal ferrimagnets},\ }\href
  {https://doi.org/10.1103/PhysRevLett.120.097202} {\bibfield  {journal}
  {\bibinfo  {journal} {Phys. Rev. Lett.}\ }\textbf {\bibinfo {volume} {120}},\
  \bibinfo {pages} {097202} (\bibinfo {year} {2018}{\natexlab{a}})}\BibitemShut
  {NoStop}%
\bibitem [{\citenamefont {Wernick}\ and\ \citenamefont
  {Geller}(1959)}]{Wernick1959}%
  \BibitemOpen
  \bibfield  {author} {\bibinfo {author} {\bibfnamefont {J.~H.}\ \bibnamefont
  {Wernick}}\ and\ \bibinfo {author} {\bibfnamefont {S.}~\bibnamefont
  {Geller}},\ }\bibfield  {title} {\bibinfo {title} {{Transition
  element{--}rare earth compounds with {C}u\textsubscript{5}Ca structure}},\
  }\href {https://doi.org/10.1107/S0365110X59001955} {\bibfield  {journal}
  {\bibinfo  {journal} {Acta Crystallogr.}\ }\textbf {\bibinfo {volume} {12}},\
  \bibinfo {pages} {662} (\bibinfo {year} {1959})}\BibitemShut {NoStop}%
\bibitem [{\citenamefont {Kumar}(1988)}]{Kumar1988}%
  \BibitemOpen
  \bibfield  {author} {\bibinfo {author} {\bibfnamefont {K.}~\bibnamefont
  {Kumar}},\ }\bibfield  {title} {\bibinfo {title} {{RETM}\textsubscript{5} and
  {RE}\textsubscript{2}{TM}\textsubscript{17} permanent magnets development},\
  }\href {https://doi.org/10.1063/1.341084} {\bibfield  {journal} {\bibinfo
  {journal} {J. Appl. Phys.}\ }\textbf {\bibinfo {volume} {63}},\ \bibinfo
  {pages} {R13} (\bibinfo {year} {1988})}\BibitemShut {NoStop}%
\bibitem [{\citenamefont {Ermolenko}(1976)}]{Ermolenko1976}%
  \BibitemOpen
  \bibfield  {author} {\bibinfo {author} {\bibfnamefont {A.}~\bibnamefont
  {Ermolenko}},\ }\bibfield  {title} {\bibinfo {title} {Magnetocrystalline
  anisotropy of rare earth intermetallics},\ }\href
  {https://doi.org/10.1109/TMAG.1976.1059178} {\bibfield  {journal} {\bibinfo
  {journal} {IEEE Trans.\ Magn.}\ }\textbf {\bibinfo {volume} {12}},\ \bibinfo
  {pages} {992} (\bibinfo {year} {1976})}\BibitemShut {NoStop}%
\bibitem [{\citenamefont {Klein}\ \emph {et~al.}(1975)\citenamefont {Klein},
  \citenamefont {Menth},\ and\ \citenamefont {Perkins}}]{Klein1975}%
  \BibitemOpen
  \bibfield  {author} {\bibinfo {author} {\bibfnamefont {H.}~\bibnamefont
  {Klein}}, \bibinfo {author} {\bibfnamefont {A.}~\bibnamefont {Menth}},\ and\
  \bibinfo {author} {\bibfnamefont {R.}~\bibnamefont {Perkins}},\ }\bibfield
  {title} {\bibinfo {title} {Magnetocrystalline anisotropy of light rare-earth
  cobalt compounds},\ }\href
  {https://doi.org/http://dx.doi.org/10.1016/0378-4363(75)90061-3} {\bibfield
  {journal} {\bibinfo  {journal} {Physica B+C}\ }\textbf {\bibinfo {volume}
  {80}},\ \bibinfo {pages} {153} (\bibinfo {year} {1975})}\BibitemShut
  {NoStop}%
\bibitem [{\citenamefont {Kuz'min}\ and\ \citenamefont
  {Tishin}(2008)}]{Kuzmin2008}%
  \BibitemOpen
  \bibfield  {author} {\bibinfo {author} {\bibfnamefont {M.~D.}\ \bibnamefont
  {Kuz'min}}\ and\ \bibinfo {author} {\bibfnamefont {A.~M.}\ \bibnamefont
  {Tishin}},\ }\bibinfo {title} {\textnormal{Theory of Crystal Field Effects in
  3$d$-4$f$ Intermetallic Compounds, in} handbook of magnetic materials},\ in\
  \href@noop {} {\emph {\bibinfo {booktitle} {Handbook of Magnetic
  Materials}}},\ Vol.~\bibinfo {volume} {17},\ \bibinfo {editor} {edited by\
  \bibinfo {editor} {\bibfnamefont {K.~H.~J.}\ \bibnamefont {Buschow}}}\
  (\bibinfo  {publisher} {Elsevier B.V.},\ \bibinfo {year} {2008})\
  Chap.~\bibinfo {chapter} {3}, p.\ \bibinfo {pages} {149}\BibitemShut
  {NoStop}%
\bibitem [{\citenamefont {Patrick}\ and\ \citenamefont
  {Staunton}(2019)}]{Patrick20192}%
  \BibitemOpen
  \bibfield  {author} {\bibinfo {author} {\bibfnamefont {C.~E.}\ \bibnamefont
  {Patrick}}\ and\ \bibinfo {author} {\bibfnamefont {J.~B.}\ \bibnamefont
  {Staunton}},\ }\bibfield  {title} {\bibinfo {title} {Crystal field
  coefficients for yttrium analogues of rare-earth/transition-metal magnets
  using density-functional theory in the projector-augmented wave formalism},\
  }\href {https://doi.org/10.1088/1361-648X/ab18f3} {\bibfield  {journal}
  {\bibinfo  {journal} {J. Phys.: Condens. Matter}\ }\textbf {\bibinfo {volume}
  {31}},\ \bibinfo {pages} {305901} (\bibinfo {year} {2019})}\BibitemShut
  {NoStop}%
\bibitem [{\citenamefont {Herbst}(1991)}]{Herbst1991}%
  \BibitemOpen
  \bibfield  {author} {\bibinfo {author} {\bibfnamefont {J.~F.}\ \bibnamefont
  {Herbst}},\ }\bibfield  {title} {\bibinfo {title}
  {R\textsubscript{2}{F}e\textsubscript{14}{B} materials: {I}ntrinsic
  properties and technological aspects},\ }\href
  {https://doi.org/10.1103/RevModPhys.63.819} {\bibfield  {journal} {\bibinfo
  {journal} {Rev. Mod. Phys.}\ }\textbf {\bibinfo {volume} {63}},\ \bibinfo
  {pages} {819} (\bibinfo {year} {1991})}\BibitemShut {NoStop}%
\bibitem [{\citenamefont {Brooks}\ \emph {et~al.}(1989)\citenamefont {Brooks},
  \citenamefont {Eriksson},\ and\ \citenamefont {Johansson}}]{Brooks1989}%
  \BibitemOpen
  \bibfield  {author} {\bibinfo {author} {\bibfnamefont {M.~S.~S.}\
  \bibnamefont {Brooks}}, \bibinfo {author} {\bibfnamefont {O.}~\bibnamefont
  {Eriksson}},\ and\ \bibinfo {author} {\bibfnamefont {B.}~\bibnamefont
  {Johansson}},\ }\bibfield  {title} {\bibinfo {title} {3$d$-5$d$ band
  magnetism in rare earth transition metal intermetallics:
  {L}u{F}e\textsubscript{2}},\ }\href
  {https://doi.org/10.1088/0953-8984/1/34/004} {\bibfield  {journal} {\bibinfo
  {journal} {J. Phys.: Condens. Matter}\ }\textbf {\bibinfo {volume} {1}},\
  \bibinfo {pages} {5861} (\bibinfo {year} {1989})}\BibitemShut {NoStop}%
\bibitem [{\citenamefont {Loewenhaupt}\ \emph {et~al.}(1994)\citenamefont
  {Loewenhaupt}, \citenamefont {Tils}, \citenamefont {Buschow},\ and\
  \citenamefont {Eccleston}}]{Loewenhaupt1994}%
  \BibitemOpen
  \bibfield  {author} {\bibinfo {author} {\bibfnamefont {M.}~\bibnamefont
  {Loewenhaupt}}, \bibinfo {author} {\bibfnamefont {P.}~\bibnamefont {Tils}},
  \bibinfo {author} {\bibfnamefont {K.}~\bibnamefont {Buschow}},\ and\ \bibinfo
  {author} {\bibfnamefont {R.}~\bibnamefont {Eccleston}},\ }\bibfield  {title}
  {\bibinfo {title} {Intersublattice exchange coupling in {G}d-{C}o compounds
  studied by {INS}},\ }\href
  {https://doi.org/http://dx.doi.org/10.1016/0304-8853(94)90398-0} {\bibfield
  {journal} {\bibinfo  {journal} {J. Magn. Magn. Mater.}\ }\textbf {\bibinfo
  {volume} {138}},\ \bibinfo {pages} {52} (\bibinfo {year} {1994})}\BibitemShut
  {NoStop}%
\bibitem [{\citenamefont {Patrick}\ \emph {et~al.}(2017)\citenamefont
  {Patrick}, \citenamefont {Kumar}, \citenamefont {Balakrishnan}, \citenamefont
  {Edwards}, \citenamefont {Lees}, \citenamefont {Mendive-Tapia}, \citenamefont
  {Petit},\ and\ \citenamefont {Staunton}}]{Patrick2017}%
  \BibitemOpen
  \bibfield  {author} {\bibinfo {author} {\bibfnamefont {C.~E.}\ \bibnamefont
  {Patrick}}, \bibinfo {author} {\bibfnamefont {S.}~\bibnamefont {Kumar}},
  \bibinfo {author} {\bibfnamefont {G.}~\bibnamefont {Balakrishnan}}, \bibinfo
  {author} {\bibfnamefont {R.~S.}\ \bibnamefont {Edwards}}, \bibinfo {author}
  {\bibfnamefont {M.~R.}\ \bibnamefont {Lees}}, \bibinfo {author}
  {\bibfnamefont {E.}~\bibnamefont {Mendive-Tapia}}, \bibinfo {author}
  {\bibfnamefont {L.}~\bibnamefont {Petit}},\ and\ \bibinfo {author}
  {\bibfnamefont {J.~B.}\ \bibnamefont {Staunton}},\ }\bibfield  {title}
  {\bibinfo {title} {Rare-earth/transition-metal magnetic interactions in
  pristine and ({N}i,{F}e)-doped {YC}o\textsubscript{5} and
  {G}d{C}o\textsubscript{5}},\ }\href
  {https://doi.org/10.1103/PhysRevMaterials.1.024411} {\bibfield  {journal}
  {\bibinfo  {journal} {Phys. Rev. Mater.}\ }\textbf {\bibinfo {volume} {1}},\
  \bibinfo {pages} {024411} (\bibinfo {year} {2017})}\BibitemShut {NoStop}%
\bibitem [{\citenamefont {Brown}(1957)}]{Brown1957}%
  \BibitemOpen
  \bibfield  {author} {\bibinfo {author} {\bibfnamefont {W.~F.}\ \bibnamefont
  {Brown}},\ }\bibfield  {title} {\bibinfo {title} {Criterion for uniform
  micromagnetization},\ }\href {https://doi.org/10.1103/PhysRev.105.1479}
  {\bibfield  {journal} {\bibinfo  {journal} {Phys. Rev.}\ }\textbf {\bibinfo
  {volume} {105}},\ \bibinfo {pages} {1479} (\bibinfo {year}
  {1957})}\BibitemShut {NoStop}%
\bibitem [{\citenamefont {Newman}\ and\ \citenamefont {Ng}(1989)}]{Newman1989}%
  \BibitemOpen
  \bibfield  {author} {\bibinfo {author} {\bibfnamefont {D.~J.}\ \bibnamefont
  {Newman}}\ and\ \bibinfo {author} {\bibfnamefont {B.}~\bibnamefont {Ng}},\
  }\bibfield  {title} {\bibinfo {title} {The superposition model of crystal
  fields},\ }\href {https://doi.org/10.1088/0034-4885/52/6/002} {\bibfield
  {journal} {\bibinfo  {journal} {Rep.\ Prog.\ Phys.}\ }\textbf {\bibinfo
  {volume} {52}},\ \bibinfo {pages} {699} (\bibinfo {year} {1989})}\BibitemShut
  {NoStop}%
\bibitem [{\citenamefont {Sucksmith}\ and\ \citenamefont
  {Thompson}(1954)}]{Sucksmith1954}%
  \BibitemOpen
  \bibfield  {author} {\bibinfo {author} {\bibfnamefont {W.}~\bibnamefont
  {Sucksmith}}\ and\ \bibinfo {author} {\bibfnamefont {J.~E.}\ \bibnamefont
  {Thompson}},\ }\bibfield  {title} {\bibinfo {title} {The magnetic anisotropy
  of cobalt},\ }\href {https://doi.org/10.1098/rspa.1954.0209} {\bibfield
  {journal} {\bibinfo  {journal} {Proc. Royal Soc. A}\ }\textbf {\bibinfo
  {volume} {225}},\ \bibinfo {pages} {362} (\bibinfo {year}
  {1954})}\BibitemShut {NoStop}%
\bibitem [{\citenamefont {Tie-song}\ \emph {et~al.}(1991)\citenamefont
  {Tie-song}, \citenamefont {Han-min}, \citenamefont {Guang-hua}, \citenamefont
  {Xiu-feng},\ and\ \citenamefont {Hong}}]{Tiesong1991}%
  \BibitemOpen
  \bibfield  {author} {\bibinfo {author} {\bibfnamefont {Z.}~\bibnamefont
  {Tie-song}}, \bibinfo {author} {\bibfnamefont {J.}~\bibnamefont {Han-min}},
  \bibinfo {author} {\bibfnamefont {G.}~\bibnamefont {Guang-hua}}, \bibinfo
  {author} {\bibfnamefont {H.}~\bibnamefont {Xiu-feng}},\ and\ \bibinfo
  {author} {\bibfnamefont {C.}~\bibnamefont {Hong}},\ }\bibfield  {title}
  {\bibinfo {title} {Magnetic properties of \textit{R} ions in
  \textit{R}{C}o\textsubscript{5} compounds (\textit{R} = {Pr, Nd, Sm, Gd, Tb,
  Dy, Ho, and Er})},\ }\href {https://doi.org/10.1103/PhysRevB.43.8593}
  {\bibfield  {journal} {\bibinfo  {journal} {Phys. Rev. B}\ }\textbf {\bibinfo
  {volume} {43}},\ \bibinfo {pages} {8593} (\bibinfo {year}
  {1991})}\BibitemShut {NoStop}%
\bibitem [{\citenamefont {Stevens}(1952)}]{Stevens1952}%
  \BibitemOpen
  \bibfield  {author} {\bibinfo {author} {\bibfnamefont {K.~W.~H.}\
  \bibnamefont {Stevens}},\ }\bibfield  {title} {\bibinfo {title} {Matrix
  elements and operator equivalents connected with the magnetic properties of
  rare earth ions},\ }\href {https://doi.org/10.1088/0370-1298/65/3/308}
  {\bibfield  {journal} {\bibinfo  {journal} {Proc. Phys. Soc. A}\ }\textbf
  {\bibinfo {volume} {65}},\ \bibinfo {pages} {209} (\bibinfo {year}
  {1952})}\BibitemShut {NoStop}%
\bibitem [{\citenamefont {Chilton}\ \emph {et~al.}(2013)\citenamefont
  {Chilton}, \citenamefont {Anderson}, \citenamefont {Turner}, \citenamefont
  {Soncini},\ and\ \citenamefont {Murray}}]{Chilton2013}%
  \BibitemOpen
  \bibfield  {author} {\bibinfo {author} {\bibfnamefont {N.~F.}\ \bibnamefont
  {Chilton}}, \bibinfo {author} {\bibfnamefont {R.~P.}\ \bibnamefont
  {Anderson}}, \bibinfo {author} {\bibfnamefont {L.~D.}\ \bibnamefont
  {Turner}}, \bibinfo {author} {\bibfnamefont {A.}~\bibnamefont {Soncini}},\
  and\ \bibinfo {author} {\bibfnamefont {K.~S.}\ \bibnamefont {Murray}},\
  }\bibfield  {title} {\bibinfo {title} {{PHI}: A powerful new program for the
  analysis of anisotropic monomeric and exchange-coupled polynuclear d- and
  f-block complexes},\ }\href {https://doi.org/10.1002/jcc.23234} {\bibfield
  {journal} {\bibinfo  {journal} {J. Comput. Chem.}\ }\textbf {\bibinfo
  {volume} {34}},\ \bibinfo {pages} {1164} (\bibinfo {year}
  {2013})}\BibitemShut {NoStop}%
\bibitem [{\citenamefont {Alameda}\ \emph {et~al.}(1981)\citenamefont
  {Alameda}, \citenamefont {Givord}, \citenamefont {Lemaire},\ and\
  \citenamefont {Lu}}]{Alameda1981}%
  \BibitemOpen
  \bibfield  {author} {\bibinfo {author} {\bibfnamefont {J.~M.}\ \bibnamefont
  {Alameda}}, \bibinfo {author} {\bibfnamefont {D.}~\bibnamefont {Givord}},
  \bibinfo {author} {\bibfnamefont {R.}~\bibnamefont {Lemaire}},\ and\ \bibinfo
  {author} {\bibfnamefont {Q.}~\bibnamefont {Lu}},\ }\bibfield  {title}
  {\bibinfo {title} {Co energy and magnetization anisotropies in {RC}o5
  intermetallics between 4.2 {K} and 300 {K}},\ }\href
  {https://doi.org/http://dx.doi.org/10.1063/1.329622} {\bibfield  {journal}
  {\bibinfo  {journal} {J.\ Appl.\ Phys.}\ }\textbf {\bibinfo {volume} {52}},\
  \bibinfo {pages} {2079} (\bibinfo {year} {1981})}\BibitemShut {NoStop}%
\bibitem [{\citenamefont {Staunton}\ \emph {et~al.}(2006)\citenamefont
  {Staunton}, \citenamefont {Szunyogh}, \citenamefont {Buruzs}, \citenamefont
  {Gyorffy}, \citenamefont {Ostanin},\ and\ \citenamefont
  {Udvardi}}]{Staunton2006}%
  \BibitemOpen
  \bibfield  {author} {\bibinfo {author} {\bibfnamefont {J.~B.}\ \bibnamefont
  {Staunton}}, \bibinfo {author} {\bibfnamefont {L.}~\bibnamefont {Szunyogh}},
  \bibinfo {author} {\bibfnamefont {A.}~\bibnamefont {Buruzs}}, \bibinfo
  {author} {\bibfnamefont {B.~L.}\ \bibnamefont {Gyorffy}}, \bibinfo {author}
  {\bibfnamefont {S.}~\bibnamefont {Ostanin}},\ and\ \bibinfo {author}
  {\bibfnamefont {L.}~\bibnamefont {Udvardi}},\ }\bibfield  {title} {\bibinfo
  {title} {Temperature dependence of magnetic anisotropy: An \textit{ab initio}
  approach},\ }\href {https://doi.org/10.1103/PhysRevB.74.144411} {\bibfield
  {journal} {\bibinfo  {journal} {Phys. Rev. B}\ }\textbf {\bibinfo {volume}
  {74}},\ \bibinfo {pages} {144411} (\bibinfo {year} {2006})}\BibitemShut
  {NoStop}%
\bibitem [{\citenamefont {Patrick}\ \emph
  {et~al.}(2018{\natexlab{b}})\citenamefont {Patrick}, \citenamefont {Kumar},
  \citenamefont {G\"otze}, \citenamefont {Pearce}, \citenamefont {Singleton},
  \citenamefont {Rowlands}, \citenamefont {Balakrishnan}, \citenamefont {Lees},
  \citenamefont {Goddard},\ and\ \citenamefont {Staunton}}]{Patrick20183}%
  \BibitemOpen
  \bibfield  {author} {\bibinfo {author} {\bibfnamefont {C.~E.}\ \bibnamefont
  {Patrick}}, \bibinfo {author} {\bibfnamefont {S.}~\bibnamefont {Kumar}},
  \bibinfo {author} {\bibfnamefont {K.}~\bibnamefont {G\"otze}}, \bibinfo
  {author} {\bibfnamefont {M.~J.}\ \bibnamefont {Pearce}}, \bibinfo {author}
  {\bibfnamefont {J.}~\bibnamefont {Singleton}}, \bibinfo {author}
  {\bibfnamefont {G.}~\bibnamefont {Rowlands}}, \bibinfo {author}
  {\bibfnamefont {G.}~\bibnamefont {Balakrishnan}}, \bibinfo {author}
  {\bibfnamefont {M.~R.}\ \bibnamefont {Lees}}, \bibinfo {author}
  {\bibfnamefont {P.~A.}\ \bibnamefont {Goddard}},\ and\ \bibinfo {author}
  {\bibfnamefont {J.~B.}\ \bibnamefont {Staunton}},\ }\bibfield  {title}
  {\bibinfo {title} {Field-induced canting of magnetic moments in
  {G}d{C}o\textsubscript{5} at finite temperature: first-principles
  calculations and high-field measurements},\ }\href
  {https://doi.org/10.1088/1361-648X/aad029} {\bibfield  {journal} {\bibinfo
  {journal} {J. Phys.: Condens. Matter}\ }\textbf {\bibinfo {volume} {30}},\
  \bibinfo {pages} {32LT01} (\bibinfo {year} {2018}{\natexlab{b}})}\BibitemShut
  {NoStop}%
\bibitem [{\citenamefont {Andreev}(1995)}]{AndreevHMM}%
  \BibitemOpen
  \bibfield  {author} {\bibinfo {author} {\bibfnamefont {A.~V.}\ \bibnamefont
  {Andreev}},\ }\bibinfo {title} {in {H}andbook of {M}agnetic {M}aterials}\
  (\bibinfo  {publisher} {Elsevier North-Holland, New York},\ \bibinfo {year}
  {1995})\ Chap.~\bibinfo {chapter} {2}, p.~\bibinfo {pages} {59}\BibitemShut
  {NoStop}%
\bibitem [{\citenamefont {Buschow}(1977)}]{Buschow1977}%
  \BibitemOpen
  \bibfield  {author} {\bibinfo {author} {\bibfnamefont {K.~H.~J.}\
  \bibnamefont {Buschow}},\ }\bibfield  {title} {\bibinfo {title}
  {Intermetallic compounds of rare-earth and 3d transition metals},\ }\href
  {https://doi.org/10.1088/0034-4885/40/10/002} {\bibfield  {journal} {\bibinfo
   {journal} {Rep.\ Prog.\ Phys.}\ }\textbf {\bibinfo {volume} {40}},\ \bibinfo
  {pages} {1179} (\bibinfo {year} {1977})}\BibitemShut {NoStop}%
\bibitem [{\citenamefont {Bruno}\ and\ \citenamefont
  {Ginatempo}(1997)}]{Bruno1997}%
  \BibitemOpen
  \bibfield  {author} {\bibinfo {author} {\bibfnamefont {E.}~\bibnamefont
  {Bruno}}\ and\ \bibinfo {author} {\bibfnamefont {B.}~\bibnamefont
  {Ginatempo}},\ }\bibfield  {title} {\bibinfo {title} {Algorithms for
  {K}orringa-{K}ohn-{R}ostoker electronic structure calculations in any
  {B}ravais lattice},\ }\href {https://doi.org/10.1103/PhysRevB.55.12946}
  {\bibfield  {journal} {\bibinfo  {journal} {Phys. Rev. B}\ }\textbf {\bibinfo
  {volume} {55}},\ \bibinfo {pages} {12946} (\bibinfo {year}
  {1997})}\BibitemShut {NoStop}%
\bibitem [{\citenamefont {Vosko}\ \emph {et~al.}(1980)\citenamefont {Vosko},
  \citenamefont {Wilk},\ and\ \citenamefont {Nusair}}]{Vosko1980}%
  \BibitemOpen
  \bibfield  {author} {\bibinfo {author} {\bibfnamefont {S.~H.}\ \bibnamefont
  {Vosko}}, \bibinfo {author} {\bibfnamefont {L.}~\bibnamefont {Wilk}},\ and\
  \bibinfo {author} {\bibfnamefont {M.}~\bibnamefont {Nusair}},\ }\bibfield
  {title} {\bibinfo {title} {Accurate spin-dependent electron liquid
  correlation energies for local spin density calculations: a critical
  analysis},\ }\href {https://doi.org/10.1139/p80-159} {\bibfield  {journal}
  {\bibinfo  {journal} {Can. J. Phys.}\ }\textbf {\bibinfo {volume} {58}},\
  \bibinfo {pages} {1200} (\bibinfo {year} {1980})}\BibitemShut {NoStop}%
\bibitem [{\citenamefont {Eriksson}\ \emph {et~al.}(1990)\citenamefont
  {Eriksson}, \citenamefont {Johansson}, \citenamefont {Albers}, \citenamefont
  {Boring},\ and\ \citenamefont {Brooks}}]{Eriksson19901}%
  \BibitemOpen
  \bibfield  {author} {\bibinfo {author} {\bibfnamefont {O.}~\bibnamefont
  {Eriksson}}, \bibinfo {author} {\bibfnamefont {B.}~\bibnamefont {Johansson}},
  \bibinfo {author} {\bibfnamefont {R.~C.}\ \bibnamefont {Albers}}, \bibinfo
  {author} {\bibfnamefont {A.~M.}\ \bibnamefont {Boring}},\ and\ \bibinfo
  {author} {\bibfnamefont {M.~S.~S.}\ \bibnamefont {Brooks}},\ }\bibfield
  {title} {\bibinfo {title} {Orbital magnetism in {F}e, {C}o, and {N}i},\
  }\href {https://doi.org/10.1103/PhysRevB.42.2707} {\bibfield  {journal}
  {\bibinfo  {journal} {Phys. Rev. B}\ }\textbf {\bibinfo {volume} {42}},\
  \bibinfo {pages} {2707} (\bibinfo {year} {1990})}\BibitemShut {NoStop}%
\bibitem [{SM()}]{SM}%
  \BibitemOpen
  \href@noop {} {}\bibinfo {note} {See Supplemental Material at [URL] for
  tables of DFT-DLM calculated parameters and CF coefficients, the experimental
  RECo\textsubscript{5} lattice constants used, a description of the
  Sucksmith-Thompson method used to extract anisotropy constants, and
  calculations testing the extent of canting in PrCo\textsubscript{5} and
  CeCo\textsubscript{5}}\BibitemShut {NoStop}%
\bibitem [{\citenamefont {Enkovaara}\ \emph {et~al.}(2010)\citenamefont
  {Enkovaara} \emph {et~al.}}]{Enkovaara2010}%
  \BibitemOpen
  \bibfield  {author} {\bibinfo {author} {\bibfnamefont {J.}~\bibnamefont
  {Enkovaara}} \emph {et~al.},\ }\bibfield  {title} {\bibinfo {title}
  {Electronic structure calculations with {GPAW}: a real-space implementation
  of the projector augmented-wave method},\ }\href
  {https://doi.org/10.1088/0953-8984/22/25/253202} {\bibfield  {journal}
  {\bibinfo  {journal} {J. Phys.: Condens. Matter}\ }\textbf {\bibinfo {volume}
  {22}},\ \bibinfo {pages} {253202} (\bibinfo {year} {2010})}\BibitemShut
  {NoStop}%
\bibitem [{\citenamefont {Elliott}(1972)}]{Elliottbook}%
  \BibitemOpen
  \bibfield  {author} {\bibinfo {author} {\bibfnamefont {R.~J.}\ \bibnamefont
  {Elliott}},\ }in\ \href@noop {} {\emph {\bibinfo {booktitle} {Magnetic
  Properties of Rare Earth Metals}}}\ (\bibinfo  {publisher} {Plenum Press},\
  \bibinfo {address} {London and New York},\ \bibinfo {year} {1972})\
  p.~\bibinfo {pages} {2}\BibitemShut {NoStop}%
\bibitem [{\citenamefont {Arrott}(1957)}]{Arrott1957}%
  \BibitemOpen
  \bibfield  {author} {\bibinfo {author} {\bibfnamefont {A.}~\bibnamefont
  {Arrott}},\ }\bibfield  {title} {\bibinfo {title} {Criterion for
  ferromagnetism from observations of magnetic isotherms},\ }\href
  {https://doi.org/10.1103/PhysRev.108.1394} {\bibfield  {journal} {\bibinfo
  {journal} {Phys. Rev.}\ }\textbf {\bibinfo {volume} {108}},\ \bibinfo {pages}
  {1394} (\bibinfo {year} {1957})}\BibitemShut {NoStop}%
\bibitem [{\citenamefont {Radwa\'nski}(1987)}]{Radwanski1987}%
  \BibitemOpen
  \bibfield  {author} {\bibinfo {author} {\bibfnamefont {R.~J.}\ \bibnamefont
  {Radwa\'nski}},\ }\bibfield  {title} {\bibinfo {title} {The origin of the
  basal-plane magnetocrystalline anisotropy in 4$f$ {C}o-rich intermetallics},\
  }\href {https://doi.org/10.1088/0305-4608/17/1/030} {\bibfield  {journal}
  {\bibinfo  {journal} {J. Phys. F: Met. Phys.}\ }\textbf {\bibinfo {volume}
  {17}},\ \bibinfo {pages} {267} (\bibinfo {year} {1987})}\BibitemShut
  {NoStop}%
\bibitem [{\citenamefont {Nikitin}\ \emph {et~al.}(2010)\citenamefont
  {Nikitin}, \citenamefont {Skokov}, \citenamefont {Koshkid'ko}, \citenamefont
  {Pastushenkov},\ and\ \citenamefont {Ivanova}}]{Nikitin2010}%
  \BibitemOpen
  \bibfield  {author} {\bibinfo {author} {\bibfnamefont {S.~A.}\ \bibnamefont
  {Nikitin}}, \bibinfo {author} {\bibfnamefont {K.~P.}\ \bibnamefont {Skokov}},
  \bibinfo {author} {\bibfnamefont {Y.~S.}\ \bibnamefont {Koshkid'ko}},
  \bibinfo {author} {\bibfnamefont {Y.~G.}\ \bibnamefont {Pastushenkov}},\ and\
  \bibinfo {author} {\bibfnamefont {T.~I.}\ \bibnamefont {Ivanova}},\
  }\bibfield  {title} {\bibinfo {title} {Giant rotating magnetocaloric effect
  in the region of spin-reorientation transition in the
  {N}d{C}o\textsubscript{5} single crystal},\ }\href
  {https://doi.org/10.1103/PhysRevLett.105.137205} {\bibfield  {journal}
  {\bibinfo  {journal} {Phys. Rev. Lett.}\ }\textbf {\bibinfo {volume} {105}},\
  \bibinfo {pages} {137205} (\bibinfo {year} {2010})}\BibitemShut {NoStop}%
\bibitem [{\citenamefont {Ohkoshi}\ \emph {et~al.}(1976)\citenamefont
  {Ohkoshi}, \citenamefont {Kobayshi}, \citenamefont {Katayama}, \citenamefont
  {Hirano}, \citenamefont {Katayama}, \citenamefont {Hirano},\ and\
  \citenamefont {Tsushima}}]{Ohkoshi1976}%
  \BibitemOpen
  \bibfield  {author} {\bibinfo {author} {\bibfnamefont {M.}~\bibnamefont
  {Ohkoshi}}, \bibinfo {author} {\bibfnamefont {H.}~\bibnamefont {Kobayshi}},
  \bibinfo {author} {\bibfnamefont {T.}~\bibnamefont {Katayama}}, \bibinfo
  {author} {\bibfnamefont {M.}~\bibnamefont {Hirano}}, \bibinfo {author}
  {\bibfnamefont {T.}~\bibnamefont {Katayama}}, \bibinfo {author}
  {\bibfnamefont {M.}~\bibnamefont {Hirano}},\ and\ \bibinfo {author}
  {\bibfnamefont {T.}~\bibnamefont {Tsushima}},\ }\bibfield  {title} {\bibinfo
  {title} {Spin reorientation in {N}d{C}o\textsubscript{5} single crystals},\
  }\href {https://doi.org/10.1063/1.30484} {\bibfield  {journal} {\bibinfo
  {journal} {AIP Conf. Proc,}\ }\textbf {\bibinfo {volume} {29}},\ \bibinfo
  {pages} {616} (\bibinfo {year} {1976})}\BibitemShut {NoStop}%
\bibitem [{\citenamefont {Kumar}\ \emph {et~al.}(2019)\citenamefont {Kumar}
  \emph {et~al.}}]{Santosh}%
  \BibitemOpen
  \bibfield  {author} {\bibinfo {author} {\bibfnamefont {S.}~\bibnamefont
  {Kumar}} \emph {et~al.},\ }\bibfield  {title} {\bibinfo {title} {Study of the
  spin reorientation transition and magnetocrystalline anisotropy in
  {N}d{C}o\textsubscript{5} using torque magnetometry},\ }\href@noop {}
  {\bibfield  {journal} {\bibinfo  {journal} {in preparation}\ } (\bibinfo
  {year} {2019})}\BibitemShut {NoStop}%
\bibitem [{\citenamefont {Herper}\ \emph {et~al.}(2017)\citenamefont {Herper},
  \citenamefont {Ahmed}, \citenamefont {Wills}, \citenamefont {Di~Marco},
  \citenamefont {Bj\"orkman}, \citenamefont {Iu\ifmmode~\mbox{\c{s}}\else
  \c{s}\fi{}an}, \citenamefont {Balatsky},\ and\ \citenamefont
  {Eriksson}}]{Herper2017}%
  \BibitemOpen
  \bibfield  {author} {\bibinfo {author} {\bibfnamefont {H.~C.}\ \bibnamefont
  {Herper}}, \bibinfo {author} {\bibfnamefont {T.}~\bibnamefont {Ahmed}},
  \bibinfo {author} {\bibfnamefont {J.~M.}\ \bibnamefont {Wills}}, \bibinfo
  {author} {\bibfnamefont {I.}~\bibnamefont {Di~Marco}}, \bibinfo {author}
  {\bibfnamefont {T.}~\bibnamefont {Bj\"orkman}}, \bibinfo {author}
  {\bibfnamefont {D.}~\bibnamefont {Iu\ifmmode~\mbox{\c{s}}\else
  \c{s}\fi{}an}}, \bibinfo {author} {\bibfnamefont {A.~V.}\ \bibnamefont
  {Balatsky}},\ and\ \bibinfo {author} {\bibfnamefont {O.}~\bibnamefont
  {Eriksson}},\ }\bibfield  {title} {\bibinfo {title} {Combining electronic
  structure and many-body theory with large databases: A method for predicting
  the nature of $4f$ states in {C}e compounds},\ }\href
  {https://doi.org/10.1103/PhysRevMaterials.1.033802} {\bibfield  {journal}
  {\bibinfo  {journal} {Phys. Rev. Materials}\ }\textbf {\bibinfo {volume}
  {1}},\ \bibinfo {pages} {033802} (\bibinfo {year} {2017})}\BibitemShut
  {NoStop}%
\bibitem [{\citenamefont {Lipp}\ \emph {et~al.}(2008)\citenamefont {Lipp},
  \citenamefont {Jackson}, \citenamefont {Cynn}, \citenamefont {Aracne},
  \citenamefont {Evans},\ and\ \citenamefont {McMahan}}]{Lipp2008}%
  \BibitemOpen
  \bibfield  {author} {\bibinfo {author} {\bibfnamefont {M.~J.}\ \bibnamefont
  {Lipp}}, \bibinfo {author} {\bibfnamefont {D.}~\bibnamefont {Jackson}},
  \bibinfo {author} {\bibfnamefont {H.}~\bibnamefont {Cynn}}, \bibinfo {author}
  {\bibfnamefont {C.}~\bibnamefont {Aracne}}, \bibinfo {author} {\bibfnamefont
  {W.~J.}\ \bibnamefont {Evans}},\ and\ \bibinfo {author} {\bibfnamefont
  {A.~K.}\ \bibnamefont {McMahan}},\ }\bibfield  {title} {\bibinfo {title}
  {Thermal signatures of the {K}ondo volume collapse in cerium},\ }\href
  {https://doi.org/10.1103/PhysRevLett.101.165703} {\bibfield  {journal}
  {\bibinfo  {journal} {Phys. Rev. Lett.}\ }\textbf {\bibinfo {volume} {101}},\
  \bibinfo {pages} {165703} (\bibinfo {year} {2008})}\BibitemShut {NoStop}%
\bibitem [{\citenamefont {Tatsumoto}\ \emph {et~al.}(1971)\citenamefont
  {Tatsumoto}, \citenamefont {Okamoto}, \citenamefont {Fujii},\ and\
  \citenamefont {Inoue}}]{Tatsumoto1971}%
  \BibitemOpen
  \bibfield  {author} {\bibinfo {author} {\bibfnamefont {E.}~\bibnamefont
  {Tatsumoto}}, \bibinfo {author} {\bibfnamefont {T.}~\bibnamefont {Okamoto}},
  \bibinfo {author} {\bibfnamefont {H.}~\bibnamefont {Fujii}},\ and\ \bibinfo
  {author} {\bibfnamefont {C.}~\bibnamefont {Inoue}},\ }\bibfield  {title}
  {\bibinfo {title} {Saturation magnetic moment and crystalline anisotropy of
  single crystals of light rare earth cobalt compounds
  {RC}o\textsubscript{5}},\ }\href {https://doi.org/10.1051/jphyscol:19711186}
  {\bibfield  {journal} {\bibinfo  {journal} {J. Phys. Colloques}\ }\textbf
  {\bibinfo {volume} {32}},\ \bibinfo {pages} {C1} (\bibinfo {year}
  {1971})}\BibitemShut {NoStop}%
\bibitem [{\citenamefont {Radwa\'nski}(1986)}]{Radwanski19862}%
  \BibitemOpen
  \bibfield  {author} {\bibinfo {author} {\bibfnamefont {R.}~\bibnamefont
  {Radwa\'nski}},\ }\bibfield  {title} {\bibinfo {title} {The rare earth
  contribution to the magnetocrystalline anisotropy in {RC}o\textsubscript{5}
  intermetallics},\ }\href
  {https://doi.org/https://doi.org/10.1016/0304-8853(86)90744-4} {\bibfield
  {journal} {\bibinfo  {journal} {J. Magn. Magn. Mater.}\ }\textbf {\bibinfo
  {volume} {62}},\ \bibinfo {pages} {120} (\bibinfo {year} {1986})}\BibitemShut
  {NoStop}%
\bibitem [{\citenamefont {Buschow}\ \emph {et~al.}(1974)\citenamefont
  {Buschow}, \citenamefont {van Diepen},\ and\ \citenamefont
  {de~Wijn}}]{Buschow1974}%
  \BibitemOpen
  \bibfield  {author} {\bibinfo {author} {\bibfnamefont {K.}~\bibnamefont
  {Buschow}}, \bibinfo {author} {\bibfnamefont {A.}~\bibnamefont {van
  Diepen}},\ and\ \bibinfo {author} {\bibfnamefont {H.}~\bibnamefont
  {de~Wijn}},\ }\bibfield  {title} {\bibinfo {title} {Crystal-field anisotropy
  of {S}m$^{3+}$ in {S}m{C}o\textsubscript{5}},\ }\href
  {https://doi.org/http://dx.doi.org/10.1016/0038-1098(74)90690-5} {\bibfield
  {journal} {\bibinfo  {journal} {Solid State Commun.}\ }\textbf {\bibinfo
  {volume} {15}},\ \bibinfo {pages} {903} (\bibinfo {year} {1974})}\BibitemShut
  {NoStop}%
\bibitem [{\citenamefont {Richter}(1998)}]{Richter1998}%
  \BibitemOpen
  \bibfield  {author} {\bibinfo {author} {\bibfnamefont {M.}~\bibnamefont
  {Richter}},\ }\bibfield  {title} {\bibinfo {title} {Band structure theory of
  magnetism in 3d-4f compounds},\ }\href
  {https://doi.org/10.1088/0022-3727/31/9/002} {\bibfield  {journal} {\bibinfo
  {journal} {J. Phys. D: Appl. Phys.}\ }\textbf {\bibinfo {volume} {31}},\
  \bibinfo {pages} {1017} (\bibinfo {year} {1998})}\BibitemShut {NoStop}%
\end{thebibliography}

%
\end{document}